\documentclass{IEEEtran}
\usepackage[utf8]{inputenc}
\usepackage[cmex10]{amsmath}
\usepackage{amsfonts,amssymb}
\usepackage{array}
\usepackage{graphicx}
\usepackage[caption=false,font=footnotesize]{subfig}
\usepackage{cite}
\usepackage{algorithm}
\usepackage{epstopdf}

\usepackage[usenames,dvipsnames]{xcolor}
\usepackage[normalem]{ulem}
\usepackage{multirow}
\newcolumntype{P}[1]{>{\centering\arraybackslash}p{#1}}
\usepackage{booktabs}
\usepackage{multicol}
\usepackage{balance}
\usepackage{hyperref} 

\usepackage{blindtext}
\usepackage{amssymb}
\usepackage{amsthm}
\usepackage{acronym} 
\usepackage{bm} 
\usepackage{soul}
\usepackage[normalem]{ulem} 
\usepackage{cancel} 

\newtheorem{remark}{Remark}
\DeclareMathOperator*{\argmax}{argmax}

\begin{document}

\acrodef{ADS-B}{Automatic Dependent Surveillance–Broadcast}
\acrodef{AIS}{Automatic Identification System}
\acrodef{AWGN}{additive white Gaussian noise}
\acrodef{CR} {code rate}
\acrodef{CRC} {cyclic redundancy check}
\acrodef{CSS}{chirp spread spectrum}
\acrodef{DtS}{direct-to-satellite}
\acrodef{ItS}{indirect-to-satellite}
\acrodef{IoT}{Internet of Things}
\acrodef{LDRO}{Low Data Rate Optimization}
\acrodef{LEO}{Low Earth Orbit}
\acrodef{LoRa}{Long Range}
\acrodef{LoRaWAN}{Long Range Wide Area Network}
\acrodef{LPWAN}{Low-Power Wide Area Network}
\acrodef{mMTC}{massive Machine-Type Communications}
\acrodef{ToA}{Time-on-Air}
\acrodef{SER}{symbol error rate}
\acrodef{SNR}{signal-to-noise ratio}


\title{Doppler Estimation and Compensation Techniques in LoRa Direct-to-Satellite Communications}

\author{Jamil~Farhat, Gianni~Pasolini, Enrico Paolini, Muhammad Asad Ullah, and Richard~Demo~Souza%
\thanks{J. Farhat is with the Federal University of Technology-Paraná (UTFPR), Curitiba, Brazil, jamilfarhat@utfpr.edu.br. G. Pasolini and E. Paolini are with CNIT-WiLab and the Department of Electrical, Electronic, and Information Engineering, University of Bologna, Bologna, Italy, gianni.pasolini@unibo.it, e.paolini@unibo.it. M. Asad Ullah is with the VTT Technical Research Centre of Finland Ltd., Espoo, Finland, asad.ullah@vtt.fi. R. D. Souza is with the Department of Electrical and Electronics Engineering, Federal University of Santa Catarina, Florianópolis, Brazil, richard.demo@ufsc.br. \\
This work has been supported in Brazil by CNPq (305021/2021-4), CAPES/STIC-AMSUD (88881.985542/2024-01) and RNP/MCTI Brasil 6G (01245.020548/2021-07).}
}

\maketitle

\begin{abstract}
Within the \ac{LPWAN} framework, the \ac{LoRa} modulation adopted by \ac{LoRaWAN} technology has garnered significant interest as a connectivity solution for \ac{IoT} applications due to its ability to offer low-cost, low-power, and long-range communications. One emerging use case of \ac{LoRa} is \ac{DtS} connectivity, which extends coverage to remote areas for supporting \ac{IoT} operations. The satellite \ac{IoT}  industry mainly prefers \ac{LEO} because it has lower launch costs and less path loss compared to Geostationary orbit. However, a major drawback of \ac{LEO} satellites is the impact of the Doppler effect caused by their mobility. Earlier studies have confirmed that the Doppler effect significantly degrades the \ac{LoRa} \ac{DtS} performance. In this paper, we propose four frameworks for Doppler estimation and compensation in \ac{LoRa} \ac{DtS} connectivity and numerically compare the performance against the ideal scenario without the Doppler effect. Furthermore, we investigate the trade-offs among these frameworks by analyzing the interplay between spreading factor, and other key parameters related to the Doppler effect. The results provide insights into how to achieve robust \ac{LoRa} configurations for \ac{DtS} connectivity.
\end{abstract}

\begin{IEEEkeywords}
Direct-to-Satellite, LEO, LoRa, LPWAN.
\end{IEEEkeywords}

\section{Introduction} \label{sec:introduction}

The \acf{IoT} is a transformative paradigm and a cornerstone of \ac{mMTC}, designed to connect a vast number of devices, typically (though not exclusively) characterized by low-power consumption and low-data-rates~\cite{Milarokostas.2023}. Its applications span diverse sectors, including industrial automation, precision agriculture, environmental monitoring, transport \& logistics, and smart cities~\cite{Arun.2024, Guo.2024}. However, each of these domains presents distinct challenges, necessitating scalable solutions that address strict constraints on power consumption, device complexity, cost-efficiency, and data rates~\cite{6g_flagship, Jouhari.2023}.

To address the need for long-range, low-power connections, both 3GPP and non-3GPP \acfp{LPWAN} have emerged as key technologies within the \ac{IoT} ecosystem, offering a trade-off between energy efficiency, coverage, data rate, and deployment cost. Their capability to provide long-range connectivity makes them particularly well-suited for large-scale \ac{IoT} deployments across diverse environments~\cite{Qin.2019, Jiang.2022, Fraire.2022, Azim.2024}.

Among non-3GPP \ac{LPWAN} technologies, \ac{LoRaWAN} has established itself as a leading connectivity solution for \ac{IoT} applications~\cite{Shanmuga.2020, Jiang.2022}. A key factor behind its success is its use of \acf{LoRa} \ac{CSS} modulation, developed and patented by Semtech~\cite{Semtech}, which enables dynamic adaptation to varying channel conditions. This flexibility allows for a trade-off between data rate and receiver sensitivity, allowing for optimized performance across a wide range of applications and coverage areas~\cite{Pasolini.2022}. In particular, \ac{LoRaWAN} is widely used for long-range communication and is increasingly considered for integration into satellite-based \ac{mMTC} networks~\cite{Doroshkin.2019}, providing a cost-effective path to global coverage~\cite{6g_flagship}.

Notably, \ac{LEO} satellites have demonstrated strong potential for supporting \ac{LoRa} \ac{DtS} communications, not only extending coverage to remote areas but also serving as a resilient backup for terrestrial networks by interconnecting \ac{LoRaWAN} gateways through satellite backhaul~\cite{Mayorga.2020,Fraire.2019,Ullah_Mikhaylov_Alves_22}.
However, for this approach to be truly effective, a significant challenge must be addressed: the high orbital velocity of \ac{LEO} satellites. To prevent falling back to Earth due to gravitational attraction, they must travel at speeds much faster than the Earth’s rotational speed. This results in relative motion between the satellite and terrestrial \ac{IoT} devices, causing the transmitted signal’s carrier frequency to vary over time, a phenomenon known as Doppler shift~\cite{Ullah.2024}. The rate at which this frequency changes, referred to as the Doppler rate, depends on the satellite's velocity, trajectory, and the relative position of the transmitter and receiver, directly impacting signal synchronization and communication reliability.

Such a dynamic behavior introduces significant difficulties, as both the Doppler shift and the Doppler rate vary over time, requiring the receiver to continuously adjust to these changes to maintain reliable communication and therefore presenting unique challenges for the physical layer of \ac{LoRa} receivers~\cite{Kodheli.2021, Centenaro.2021}. It is worth noting, in this regard, that the Doppler shift primarily affects the beginning and end of the satellite's visibility window, causing significant variations in the carrier frequency at these points~\cite{Ullah.2024}. In contrast, the Doppler rate has the most noticeable impact during the middle of the visibility window~\cite{Ullah.2024}. The combined effect of both phenomena leads to packet losses which further reduces the effective visibility duration. 

This work aims to extend the effective communication window by mitigating Doppler-related impairments, thereby increasing the number of available transmission opportunities for IoT devices. As a result, it enhances not only the performance of the modulation scheme but also the overall efficiency of the network.

\subsection{Novelty and Contributions}

In the context of satellite communications, there are two primary access modes: \ac{ItS} and \acf{DtS}. In the \ac{ItS} mode, an intermediary terrestrial gateway is required to facilitate communication between the satellite and the terrestrial \ac{IoT} device. This intermediary gateway typically acts as a relay, forwarding the data between the \ac{IoT} device and the satellite. Conversely, \ac{DtS} mode allows \ac{IoT} devices to communicate directly with satellites, removing the dependency on ground-based infrastructure and thereby simplifying network deployment~\cite{Fraire.2019, Alvarez.2022}. This makes \ac{DtS} a preferred connectivity solution for achieving global coverage.

In this paper, we present and discuss strategies for estimating and compensating for Doppler-related impairments at the satellite, with the goal of enhancing the performance of \ac{LoRa} \ac{DtS} systems. Specifically, our contributions consist of four strategies, each utilizing specific chirps within the \ac{LoRa} frame as pilots to estimate the Doppler effect. These solutions potentially offer increasing accuracy, each addressing the challenge through distinct methods: \textit{i)} \emph{point} estimation, which uses the last downchirp in the \ac{LoRa} frame structure to estimate the Doppler shift; \textit{ii)} \emph{linear} estimation, which also calculates the angular coefficient of Doppler shift variations over time by comparing the first and last  downchirps within the \ac{LoRa} frame; and \textit{iii)} \emph{midamble} based estimations, which build upon the linear and point estimation by refining them at specified intervals, further improving accuracy.

It is important to emphasize that these solutions require no complex computations, and their implementation relies solely on software/firmware updates. Each strategy is carefully evaluated using numerical simulations that examine their performance under different Doppler scenarios. The results highlight the trade-offs between error rates and compensation accuracy, providing valuable insights into their feasibility in real-world deployments of \ac{LoRa} \ac{DtS} systems. In summary, the main contributions of this work are as follows: 
\begin{itemize} 
    \item We propose and evaluate four Doppler estimation approaches, point, linear, midamble-point, and midamble-linear, and analyze their effectiveness in reducing error rates and improving compensation accuracy. 
    \item We compare the trade-offs between these approaches through detailed numerical simulations, highlighting their relative strengths and weaknesses in mitigating Doppler effects in \ac{LoRa} \ac{DtS} communication systems.   
    \item We demonstrate, by numerical results, that the midamble-point-based estimation approach delivers consistent performance, performing nearly as well as the best approach under various conditions, including different SNR levels, payload lengths, and satellite positions. Additionally, we analyze the optimal midamble-point update interval to enhance error performance metrics.
    \item Our results underscore the critical importance of selecting appropriate Doppler estimation strategies based on key \ac{LoRa} parameters, such as the spreading factor and the interplay between Doppler shift, signal configuration, and system constraints.
\end{itemize}

The remainder of this paper is organized as follows. Section~\ref{sec:SotA} presents the state-of-the-art in related research. Section~\ref{sec:preliminaries} provides an overview of the key characteristics of \ac{LoRa} signals, including the frame structure, while \ref{sec:doppler_effects} introduces fundamental concepts related to Doppler effects. Following this, Section~\ref{sec:impact_doppler} examines the impact of Doppler effects on \ac{LoRa} \ac{DtS} communications. Section~\ref{sec:proposed_frameworks} introduces the proposed frameworks for Doppler estimation, outlining strategies to mitigate Doppler shift and rate in satellite-based systems. Section~\ref{sec:numerical_results} presents the numerical results obtained from simulations, and Section~\ref{sec:final_comments} concludes the paper with final remarks.

\section{State-of-the-Art} 
\label{sec:SotA}

\subsection{Link Budget and Scalability Aspects}

Recent studies have extensively analyzed the performance of \ac{LoRa} in various contexts, primarily focusing on link budget and scalability aspects~\cite{Nguyen.2019, Pasolini.2022, Maleki.2024, Alvarez.2022, Fernandez.2020, Zhang.2024, Alonso.2024, Wu.2024, Testi.2025}. For instance,~\cite{Nguyen.2019} proposes novel modulation and demodulation techniques to enhance efficiency and performance, with insights into \ac{LoRa} signal characteristics and demodulation techniques provided in~\cite{Chiani.2019,Pasolini.2022}. More recently, a comprehensive overview of CSS modulation, including signal generation, detection, and error performance, is presented in~\cite{Maleki.2024}, offering a deeper understanding of \ac{LoRa}'s capabilities in IoT networks.

In the context of \ac{LoRa} \ac{DtS} networks, advancements in scalability and concurrent uplink transmissions have been explored. For example,~\cite{Zhang.2024} introduces a fractional chirp rate-based CSS modulation scheme to enable non-orthogonal multiple access while maintaining noise immunity. Similarly,~\cite{Alonso.2024} proposes an ALOHA-based scheme for overloaded \ac{LoRa}-DtS networks, where devices dynamically adjust transmission rates to optimize performance. Additionally,~\cite{Wu.2024} presents an efficient random access scheme with a preamble structure designed for joint estimation of timing advance and carrier frequency offset, though limited to integer subcarrier spacing shifts. Finally,~\cite{Testi.2025} investigates collision-free uplink transmissions in uncoordinated LEO satellite IoT networks. 
Although these studies collectively demonstrate the feasibility of \ac{LoRa} DtS, the Doppler effects remain an open challenge. In particular, while significant progress has been made in improving the link budget and scalability, further research is needed  on how to estimate and compensate for Doppler shift and Doppler rate in LEO satellite communications.

\subsection{Doppler Aspects}
Doppler effects in \ac{LoRa}-based satellite communication have also been widely studied through field trials, laboratory experiments, real-world flight tests, and analytical models. In~\cite{Colombo.2022}, an experimental setup is presented where a software-defined radio (SDR) emulates Doppler effects on a \ac{LoRa} satellite link operating at $868$~MHz. The study concludes that \ac{LoRa} technology exhibits high immunity to static Doppler effects, whereas dynamic Doppler effects introduce a higher error rate.

Regarding laboratory and outdoor experiments involving device mobility, the study in~\cite{kmdoppler} investigates a \ac{LoRa} link in which the end device is mounted on a lathe machine rotating at $1500$~rounds per minute, corresponding to an equivalent linear velocity of approximately $80$~km/h. In~\cite{V2X}, a \ac{LoRa} end device was instead installed on a vehicle moving at variable speeds ranging from $32$ to $104$~km/h, causing Doppler shifts between $300$ and $400$~Hz. Both experiments confirm that \ac{LoRa} is sensitive to the Doppler effect. This is particularly significant because the Doppler effect caused by vehicular mobility is much lower than that encountered in typical \ac{LEO} satellite communications, where satellites orbit at approximately $27,000$~km/h.

In~\cite{Doroshkin.2019,LoRaDoppler1} laboratory experiments considered a \ac{LEO} satellite orbiting at altitudes of $200$~km and $550$~km, operating in the $430$~MHz-$434$~MHz frequency band. Following these experiments, these works resulted in the development of \ac{LoRa} satellite named Norby, which was launched on September 28, 2020. Norby carries an on-board SX1278 transceiver and operates at a carrier frequency of $436.703$~MHz. At the time of writing this paper, approximately one million \ac{LoRa} packets from Norby have been successfully received by multiple terrestrial gateways. Moreover, over 21 million telemetry packets from various \ac{LoRa} satellites have been successful received by terrestrial \ac{LoRaWAN} gateways. 
This impressive milestone confirms the feasibility of \ac{LoRa} \ac{DtS} communications~\cite{tinygs}. Moreover, the Norby mission provided valuable insights into the limitations of \ac{CSS} modulation in \ac{LEO} satellite communications. Particularly, the initial Norby flight-test experiments, described in~\cite{Norby}, helped understanding the limitations of \ac{LoRa} modulation at $436.703$~MHz and an orbital altitude of $560$~km.   These experiments  used a variety of configurations, including bandwidths $B\in \{ 31.25, 62.5, 125, 500\}$~kHz and payload sizes of $55$ and $143$~bytes. With such parameters, twenty experiments, each corresponding to a unique configuration, were conducted to investigate the impact of Doppler effect on \ac{LoRa} robustness. 

Similar to terrestrial \ac{LoRa} studies~\cite{kmdoppler,V2X} and \ac{LoRa} \ac{DtS} laboratory experiments~\cite{Doroshkin.2019,LoRaDoppler1}, the Norby experiments described in~\cite{Norby} confirm \ac{LoRa}'s immunity to the Doppler effect only for a limited set of configurations. Additionally, the flight-test experiments provided valuable insights into the distinct roles played by Doppler shift and Doppler rate in affecting link reliability and demodulation performance. Particularly, these experiments led to the following key findings: 
\begin{itemize}
		\item the Doppler shift degrades \ac{LoRa} signal reception and causes communication failure at the lowest elevation angles, which correspond to the maximum link distance and the highest Doppler shift;
		\item the Doppler rate challenges the receiver’s frequency tracking loop and causes communication failure  at high elevation angles, which typically corresponds to minimum link distance, where the Doppler rate reaches its highest value (in absolute terms).
\end{itemize}

As examined in~\cite{Ullah.2024,Norby}, at a $90$~degree elevation angle, the Doppler shift is almost $0$~kHz, presenting no challenges to successful signal reception. However, at the same $90$~degree elevation, there is a very high Doppler rate which challenges the \ac{LoRa} packets with high time-on-air. This makes \ac{LoRa} modulation particularly vulnerable when configurations that increase the transmission duration are used.

To mitigate Doppler-induced impairments, \cite{Colavolpe.2019} proposes a receiver architecture that integrates frequency correction mechanisms, leveraging Doppler estimation, followed by compensation, based on the preamble symbols in \ac{LoRa} frames. This notable work was a source of inspiration for our study, prompting further investigation. In fact, \cite{Colavolpe.2019} does not explore the potential impact of the built-in \ac{LDRO} mode available in \ac{LoRa} receivers, which, as highlighted in this paper, may offer advantages under high Doppler conditions. Additionally, the study does not include a comparison with alternative Doppler estimation/compensation strategies, limiting the evaluation of the proposed solution's relative effectiveness. A similar solution is proposed in \cite{sym14040747}, which also estimates frequency offset in \ac{LoRa} communications based on preamble symbols. However, this work does not address satellite links and does not specifically consider the Doppler effect and its unique characteristics. 

A different approach is presented in \cite{9943752}, where Doppler estimation, followed by compensation, is performed using pilot symbols embedded in the payload of \ac{LoRa} frames. However, the effectiveness of this strategy is evaluated using only a single figure in the numerical results, which prevents a comprehensive assessment of its performance across varying modulation parameters, pilot symbol settings, Doppler conditions, and scenarios where \ac{LDRO} is enabled.  Moreover, as in the previous case, no comparison is provided with alternative Doppler estimation/compensation strategies.

In this paper, we introduce four Doppler estimation strategies that leverage either preamble symbols or pilot symbols embedded in the payload. These approaches employ different techniques to estimate both the Doppler shift and Doppler rate over the duration of the frame. Their effectiveness is evaluated under varying conditions, including different modulation parameters, satellite orbital positions, and with or without the \ac{LDRO} mode enabled.

\section{Preliminaries on LoRa and its Adoption in \ac{DtS} Communications} \label{sec:preliminaries}

\subsection{LoRa Modulation} \label{ssec:lora_signal}

\ac{LoRa} is a specialized implementation of the \ac{CSS} modulation, designed to enable low-power communication, provide high immunity to interference, and facilitate time and frequency synchronization at the receiver. It is based on a set of linear chirps---short-time sinusoids whose frequencies sweep linearly over a given bandwidth $B$~\cite{Pasolini.2022, Ullah.2024}---making it a particular case of frequency modulation with linear frequency deviation. Specifically, \ac{LoRa} employs $M$ uniquely shaped chirps, each representing a distinct symbol $s$ in the modulation alphabet ${\mathcal{S} = \{0, 1, \dots, M-1\}}$.

From a mathematical viewpoint, a chirp can be expressed as~\cite{Pasolini.2022}
\begin{equation}
    c(s, t) = V_0 \cos\left(2 \pi F_\mathrm{C} t + 2 \pi \int_0^{t} \Delta f(s, \xi) \, \mathrm{d} \xi \right), \quad 0 \leq t \leq T_\text{c},
    \label{eq:chirp_equation}
\end{equation}
where $V_0$ represents the amplitude, $F_\mathrm{C}$ is the center frequency of the sweep interval $\left[F_\mathrm{C} - \frac{B}{2}, F_\mathrm{C} + \frac{B}{2}\right]$,  $T_\text{c}$ is the chirp duration, and $\Delta f(s, t)$ denotes the instantaneous frequency deviation from $F_\mathrm{C}$, which depends on $s$ and varies within $\left[-\frac{B}{2}, \frac{B}{2}\right]$. 

\begin{figure}[!t] 
\centering
\includegraphics[width=8.8cm]{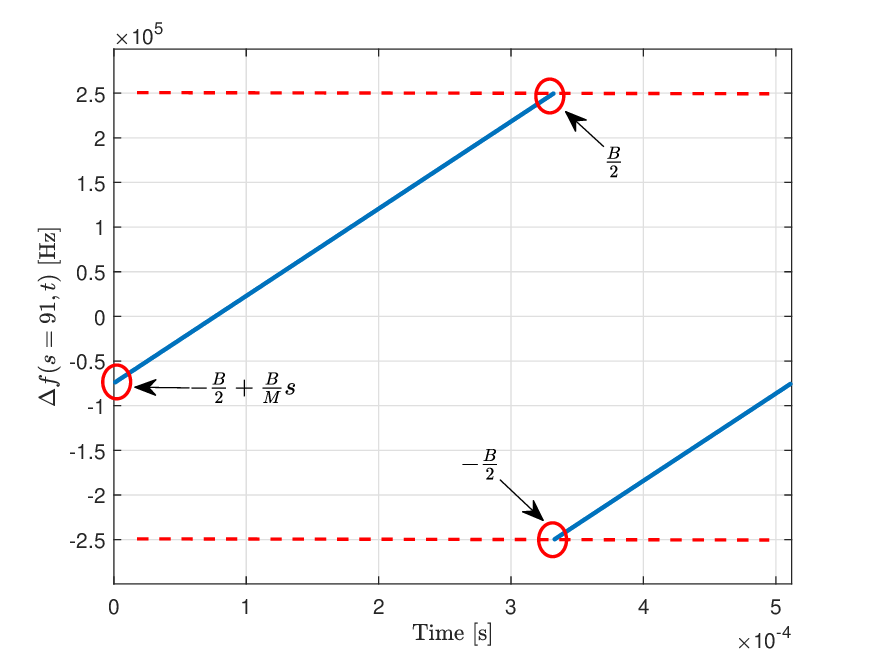}
\caption{Representation of the instantaneous frequency deviation. $s=91$, ${B=500}$~kHz, ${T_\text{c}=5.12\cdot10^{-4}}$~s, with the frequency boundaries highlighted by the dashed red lines.}
\label{fig:chirp}
\end{figure}

More precisely, for each symbol $s\in\mathcal{S}$, the instantaneous frequency deviation $\Delta f(s, t)$ of the corresponding chirp starts at ${-\frac{B}{2} + \frac{B}{M}s}$ and increases linearly to $\frac{B}{2}$. Upon reaching the maximum offset $\frac{B}{2}$, $\Delta f(s, t)$ wraps around to $-\frac{B}{2}$ and continues to increase linearly until the initial value  ${-\frac{B}{2} + \frac{B}{M}s}$ is reached at $t=T_\text{c}$. Fig.~\ref{fig:chirp} illustrates an example of instantaneous frequency deviation with $M=256$, $s=91$, $B=500$~kHz, and ${T_\text{c}=5.12\cdot10^{-4}}$s.

Remarkably, \ac{LoRa} enables a straightforward trade-off between sensitivity and data rate. This balance is achieved by carefully selecting two key parameters: the bandwidth $B$ and the spreading factor $\mathrm{SF}$. Specifically, increasing the bandwidth---configurable to $125$~kHz, $250$~kHz, or $500$~kHz---enhances the bit rate at the expense of reduced sensitivity. The spreading factor, in turn, ranges from $\mathrm{SF}=7$ to $\mathrm{SF}=12$ and influences the chirp duration, ultimately impacting communication reliability.  In fact, \ac{LoRa} is designed to adhere to the  relationship $B\,T_\text{c} = M$, with $M=2^{\mathrm{SF}}$.
Therefore, for a given bandwidth $B$, increasing  $\mathrm{SF}$ leads to a longer chirp duration, thereby enhancing communication  reliability.

This interplay between bandwidth, spreading factor, and symbol duration ensures that \ac{LoRa} can adapt to different application requirements, establishing it as a successful technology.

\subsection{LoRa Demodulation}
\label{Sec:LoRa_demodulation}
To provide a clearer understanding of the Doppler estimation strategies introduced in the following sections, we first revisit the \ac{LoRa} demodulation process, as outlined in~\cite{Pasolini.2022}. Based on~\eqref{eq:chirp_equation}, and assuming that in the context of payload demodulation $t=0$~s represents the start of the payload containing $n_{\text{sym}}$ chirps, the chirp corresponding to the $n$-th data symbol $s_n$ can be expressed as
\begin{equation}
    c_n(t) = V_0 \cos\left[2 \pi F_\mathrm{C} t + \omega(s_n,t)\right], \quad n T_\text{c} \leq t \leq (n+1) T_\text{c}
\end{equation}
where $n = \left\{0, 1, \dots, n_{\text{sym}} -1 \right\}$, and $\omega(s_n,t)$ is the phase deviation given by
\begin{equation}
    \omega(s_n, t) = 2 \pi \int_{nT_\text{c}}^{t} \Delta f(s_n, \xi) \mathrm{d} \xi,
\end{equation}
with $\Delta f(s_n, t)$ representing the instantaneous frequency deviation corresponding to $s_n$ and $nT_\text{c}\leq t \leq (n+1)T_\text{c}$. Consequently, the corresponding complex envelope within the $n$-th  symbol interval is given by
\begin{equation}
    i_n(t) = V_0 \exp\left[j \omega(s_n,t)\right].
    \label{eq:complex_envelope}
\end{equation}  

Given \eqref{eq:complex_envelope}, a dechirping operation is performed at the receiver by multiplying $i_n(t)$ with the complex conjugate of a pure upchirp\footnote{The complex conjugate of a pure upchirp is a pure downchirp, whose frequency deviation decreases linearly from $\frac{B}{2}$ to $-\frac{B}{2}$.},  yielding 
\begin{equation}
    r_n(t) = i_n(t) \exp\left[-j \theta(t - n T_\text{c})\right],
    \label{eq:dechirped_signal}
\end{equation}  
where the phase deviation $\theta(t)$ of a pure upchirp is given by
\begin{equation}
    \theta(t) = 2 \pi \left[-\frac{B}{2} t + \frac{B}{2 T_\text{c}} t^2 \right], \quad 0 \leq t \leq T_\text{c}.
\end{equation}
It is important to note that dechirping requires frame synchronization, which can still be achieved even in the presence of Doppler effects~\cite{Colavolpe.2019}. 

Demodulation is then completed by computing the Fourier transform of~\eqref{eq:dechirped_signal} and extracting the transmitted symbol $s_n$ from the frequency bin, numbered from 0 to $M-1$, where the spectrum peak appears. Unfortunately, in the presence of Doppler effect, a frequency shift is introduced, which may lead to symbol misalignment. This highlights the need for effective strategies to estimate and compensate for such frequency deviations, which are the primary focus of this paper. To achieve this, we leverage the structure of the \ac{LoRa} frame, which is detailed in the following section.

\subsection{LoRa Frame Structure} \label{ssec:lora_frame}

\begin{figure}[!t] 
\centering
\includegraphics[width=8.8cm]{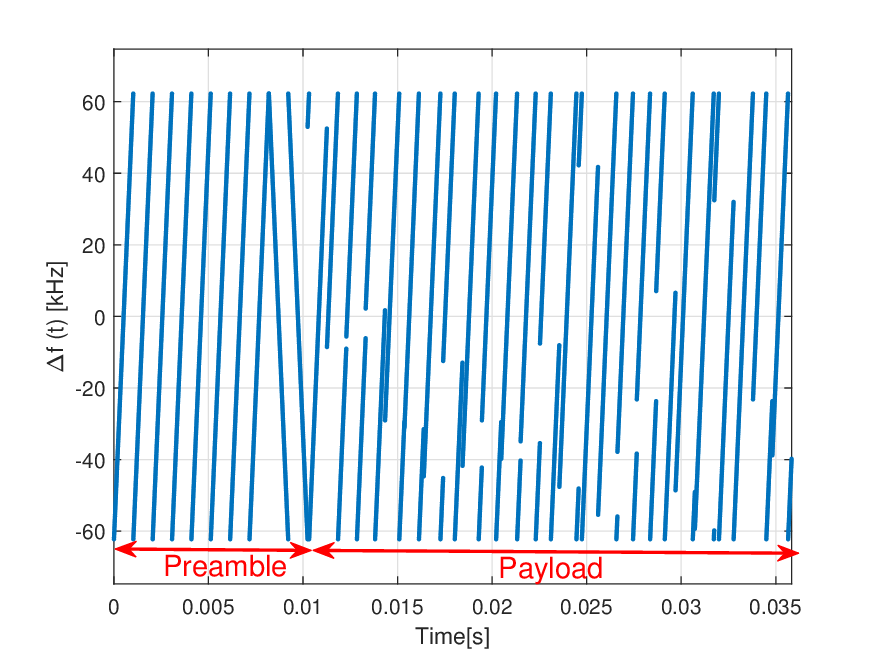}
\caption{Frequency deviation within a LoRa frame. ${B=125}$~kHz, SF=7, ${T_\text{c}}$=1~ms.}
\label{fig:frame_format}
\end{figure}

The \ac{LoRa} frame format  consists of several components: a preamble, a frame header (optional), the payload, and optionally a \ac{CRC} to enable error detection. An example is shown in Fig.~\ref{fig:frame_format}, which illustrates the frequency deviation for a frame consisting of the preamble and the payload, with ${B = 125}$~kHz and $\mathrm{SF} = 7$.

The preamble is crucial for synchronizing the receiver with the transmitted signal and mainly consists of $n_{\text{up}}$ pure upchirps. For instance, Fig.~\ref{fig:frame_format} illustrates a preamble containing $n_{\text{up}}=8$~ upchirps at the beginning of the frame. Each upchirp has a frequency deviation that increases linearly from $-\frac{B}{2}$ to $\frac{B}{2}$. This sequence of upchirps is followed by $n_{\text{dw}}$ downchirps, typically set to $n_{\text{dw}} = 2.25$, marking the end of the preamble and ensuring proper frame synchronization at the receiver\footnote{Fig.~\ref{fig:frame_format} shows only two downchirps, omitting the incomplete downchirp, as it is not relevant to the discussion in this paper.}. As it is evident in Fig.~\ref{fig:frame_format}, the frequency deviation of the downchirps decreases linearly, reversing the sweep of the upchirp. The remainder of the frame consists of the payload, which carries the actual data to be transmitted. It is composed of $n_{\text{sym}}$ chirps, each representing one symbol from the modulation alphabet ${\mathcal{S}}$, which encodes $M=2^{\mathrm{SF}}$ bits. For example, the payload in the frame shown in Fig.~\ref{fig:frame_format} consists of 25 chirps, starting immediately after the preamble downchirps. As discussed in Sec.~\ref{ssec:lora_signal}, the initial frequency of each chirp in the payload depends on the modulation symbol ${s \in \mathcal{S}}$ transmitted during the corresponding chirp interval.

In \ac{LoRa} terminology, the duration of a frame transmission is referred to as the \ac{ToA}. It is important to note that the \ac{ToA} depends not only on the number of symbols (i.e., chirps) in the payload, but also on the $\mathrm{SF}$, which determines the duration of each chirp. This is particularly relevant because longer frames, typically associated with higher $\mathrm{SFs}$, are more susceptible to the Doppler rate, which can introduce significant frequency variations between the start and end of a frame.

\subsection{Potential LoRa DtS Applications} 
\begin{table}
\caption{Potential satellite IoT applications and traffic models.}
\label{tab:payload}
\resizebox{\columnwidth}{!}{\begin{tabular}{@{}lccccc@{}}
\toprule
Application     & Report period   &Payload   &\multicolumn{3}{c}{ToA [ms]}\\
  & seconds & bytes & $\mathrm{SF}=7$ &$\mathrm{SF}=10$&$\mathrm{SF}=12$\\
\midrule
Air traffic (ADS-B)&1 & 32 &92.4&575.5&2138.1\\
Marine traffic (AIS) & 2-180  &21 &77.1&452.6&1810.4\\
Vehicle tracking &120 &8 &56.6&370.7&1482.8\\
Wildlife tracking &120 &10 &61.7&370.7&1482.8\\
Wind turbine &600  &10 &61.7&370.7&1482.8\\
Smart meter  &900 &12 &61.7&411.6&1482.8\\
Smart agriculture &1800 &10 &61.7&370.7&1482.8\\
\hline
\end{tabular}}
\end{table}
Table~\ref{tab:payload} shows \ac{DtS} IoT data traffic models including payload size, reporting period, and \ac{ToA}, considering a bandwidth of $125$~kHz, for different applications. For instance, aircraft \ac{ADS-B} systems transmit $32$-byte messages every second. \ac{AIS} transmits $21$-byte messages every 2 seconds for ships traveling over $23$~knots~\cite{Asad23,Asad_ships}, while wind turbine monitoring systems send $10$-byte payloads every $600$~seconds~\cite{Ullah_Mikhaylov_Alves_22}. In smart agriculture, devices typically transmit $10$~bytes every $1800$~seconds~\cite{Lin}, and smart meters report $12$-byte packets at regular intervals. As anticipated in Sec.~\ref{ssec:lora_frame}, the varying transmission intervals across different applications result in different Doppler effects experienced within each frame.

\section{Doppler Effect} \label{sec:doppler_effects}

\begin{figure}[!t]
    \centering
        \includegraphics[width=8.8cm]{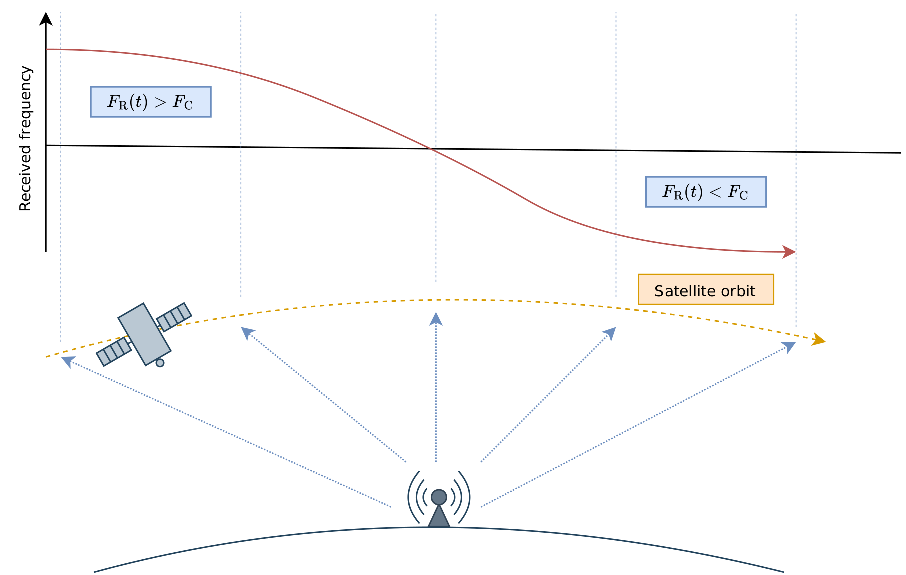}
        \caption{Variation of the received carrier frequency, $F_{\mathrm{R}}(t)$, over time due to the satellite's motion along its orbit.}
    \label{fig:DopplerRepresentation}
\end{figure}

As previously mentioned, the the relative velocity of a terrestrial end device and a \ac{LEO} satellite introduces significant challenges for \ac{DtS} communications. To maintain their orbits, \ac{LEO} satellites move at speeds much higher than the Earth's rotational velocity, resulting in substantial relative velocity with terrestrial \ac{IoT} devices. As illustrated in Fig.~\ref{fig:DopplerRepresentation}, this relative velocity causes an offset between the transmitted carrier frequency $F_{\mathrm{C}}$ and the received carrier frequency $F_{\mathrm{R}}(t)$. This phenomenon, called Doppler effect,  significantly impacts the physical layer of receivers, posing challenges in signal acquisition, synchronization, and demodulation\cite{Kodheli.2021}.

Specifically, the Doppler effect consists of two key phenomena: Doppler shift, which refers to the instantaneous frequency change due to relative motion, and Doppler rate, which describes how quickly this frequency shift varies over time. In an ideal end device and satellite geometry, the Doppler shift is most pronounced at low elevation angles, where the distance between the satellite and the \ac{IoT} device is higher. Conversely, the Doppler rate becomes more significant at high elevation angles, where the satellite is nearly overhead and the link distance is minimal. 

Mathematically, the Doppler shift $F_{\mathrm{D}}(t)$ at any given time $t$ is defined as~\cite{Ullah.2024}
\begin{equation}
F_\mathrm{D}(t) = F_\mathrm{R}(t) - F_\mathrm{C},
\end{equation}
where $F_{\mathrm{R}}(t)$ is the received carrier frequency, which varies over time due to the satellite's motion, and $F_{\mathrm{C}}$ is the transmitted carrier frequency. The Doppler shift experienced by the satellite is given by
\begin{equation}
	F_\mathrm{D}(t) = -\frac{v(t)}{c}F_\mathrm{C},
    \label{eq5}
\end{equation}
where $v(t)$ is the relative velocity\footnote{$v(t)$ is assumed to be negative, thus generating a positive Doppler shift, when the satellite is moving toward the \ac{IoT} device. In contrast, if the satellite is moving away from the \ac{IoT} device, $v(t)$ will be positive, resulting in a negative Doppler shift.} of the satellite as seen by an \ac{IoT} device~\footnote{In real scenarios, computing the relative velocity $v(t)$ requires accurate satellite data (position and velocity), precise IoT device location (latitude, longitude, altitude), and time synchronization to correlate received signals with the satellite's position. Obtaining precise data can be challenging, as it often relies on satellite operators or tracking services. Additionally, errors in device positioning or timing can significantly impact the accuracy of $v(t)$.}, and $c$ denotes the speed of light\footnote{Equation \eqref{eq5} provides a good approximation of the Doppler shift when $v(t)$ is small compared to $c$, which holds true in all practical cases.}.

On the other hand, the Doppler rate,  $\Delta F_{\mathrm{D}}(t)$, is mathematically defined as the time derivative of the Doppler shift:
\begin{equation}
    \Delta F_{\mathrm{D}}(t) = \frac{\mathrm{d}F_{\mathrm{D}}(t)}{\mathrm{d}t}.
\end{equation}

\begin{figure}[!t]
    \centering
        \includegraphics[width=8.8cm]{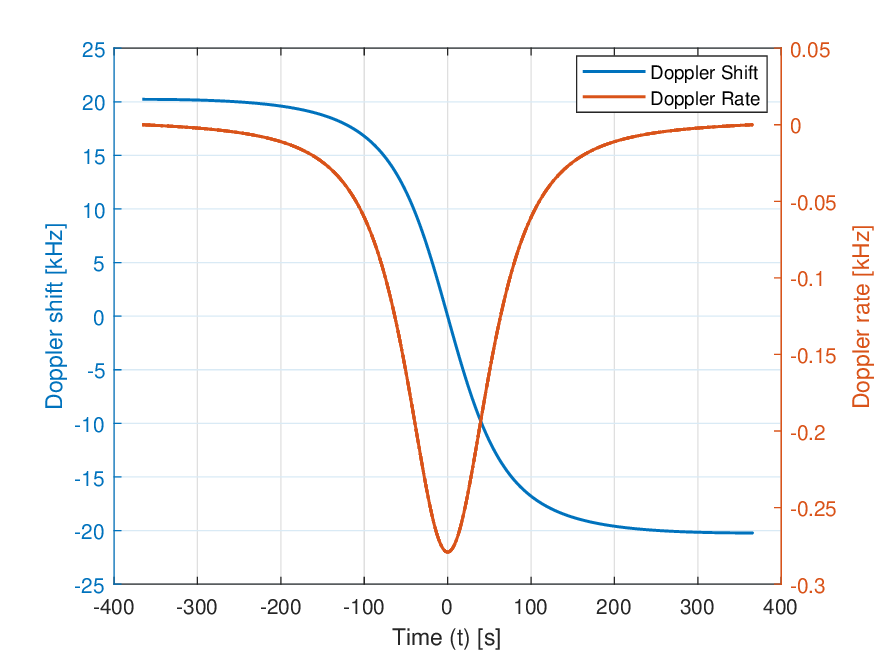}
        \caption{The \ac{LEO}-satellite Doppler Shift and Doppler Rate for an orbital distance of 550~km at a carrier frequency of 868~MHz.}
    \label{fig:DopplerEffects}
\end{figure}

Clearly, both  Doppler shift and Doppler rate are highly dependent on the satellite's position, velocity, trajectory as well as on the carrier frequency. As an example, the magnitude of both phenomena, derived from the theoretical framework outlined in~\cite{Doroshkin.2019}, are illustrated in Fig.~\ref{fig:DopplerEffects}, considering ${F_\mathrm{C}=868}$~MHz and a satellite orbital height of $550$~km \cite{Semtech.AppNote_Doppler}. In this figure, ${t = 0}$~s corresponds to the instant when the satellite is directly above the \ac{IoT} device (i.e., at the highest elevation angle), while ${t=\pm 366}$~s marks the points of lowest elevation angle. As expected, a more pronounced Doppler shift occurs at low elevation angles, whereas the highest Doppler rate is observed at the highest elevation angle.

\section{Impact of Doppler on LoRa communications}
\label{sec:impact_doppler}

Both Doppler shift and Doppler rate can severely degrade \ac{LoRa} signal reception, especially when they exceed the receiver's tolerance thresholds. For instance, a high Doppler shift may prevent the receiver from locking onto the carrier frequency, while a high Doppler rate can disrupt carrier synchronization even after it has been established, ultimately leading to frame loss.

Notably, \ac{LoRa} modulation incorporates a built-in strategy to enhance resistance to Doppler effects through the \ac{LDRO} mode. For a given $\mathrm{SF}$, this mode reduces the number of bits transmitted per symbol, effectively decreasing the modulation alphabet's cardinality to ${M = 2^{\mathrm{SF}-2}}$. By lowering symbol density, the frequency separation $\frac{B}{M}$ between chirps increases by a factor of four, thus improving robustness against Doppler\footnote{It is important to note that \ac{LDRO} mode is mandatory for higher spreading factors (e.g., $\mathrm{SF}=11$ and $\mathrm{SF}=12$) when operating at a 125 kHz bandwidth. This requirement arises because the extended frame duration at higher spreading factors makes the system more susceptible to Doppler~\cite{SX1272}.}. However, this increased robustness comes at the cost of reduced data rates, as fewer bits are mapped to a symbol.

As anticipated in Sec.~\ref{ssec:lora_signal}, $\mathrm{SF}$ is a key parameter in \ac{LoRa} communication, as it directly influences receiver sensitivity and higher $\mathrm{SF}$ values allow for extended transmission distances between the transmitter and receiver. This characteristic is particularly crucial in \ac{DtS} communication, where long-range connectivity is essential, making higher $\textrm{SFs}$ especially beneficial for enhancing receiver sensitivity.

However, the choice of $\mathrm{SF}$ also affects the system's response to the Doppler effect. In particular, as $\mathrm{SF}$ increases, the frequency separation between chirps decreases, making the receiver more susceptible to Doppler shift. Additionally, the Doppler shift experienced at the beginning of a received frame differs from that at the end, with its variation depending not only on $\Delta F_{\mathrm{D}}(t)$ but also on the frame duration, i.e., the \ac{ToA}. Since in \ac{LoRa} communication the \ac{ToA} increases with $\mathrm{SF}$, configurations with higher $\mathrm{SF}$ are more susceptible to the Doppler effect, as the carrier frequency may change significantly during the reception of the frame. It turns out, therefore, that $\textrm{SF}$ impacts receiver sensitivity and Doppler robustness in opposing ways, highlighting the importance of accurately estimating and compensating for the Doppler effect to ensure reliable communication in \ac{DtS} scenarios.

\begin{figure}[!t]
    \centering
    \subfloat[\label{fig:dopplerScenario1} \textit{Case 1}]{%
        \includegraphics[trim= {3 0 30 20}, clip, width=0.5\linewidth]{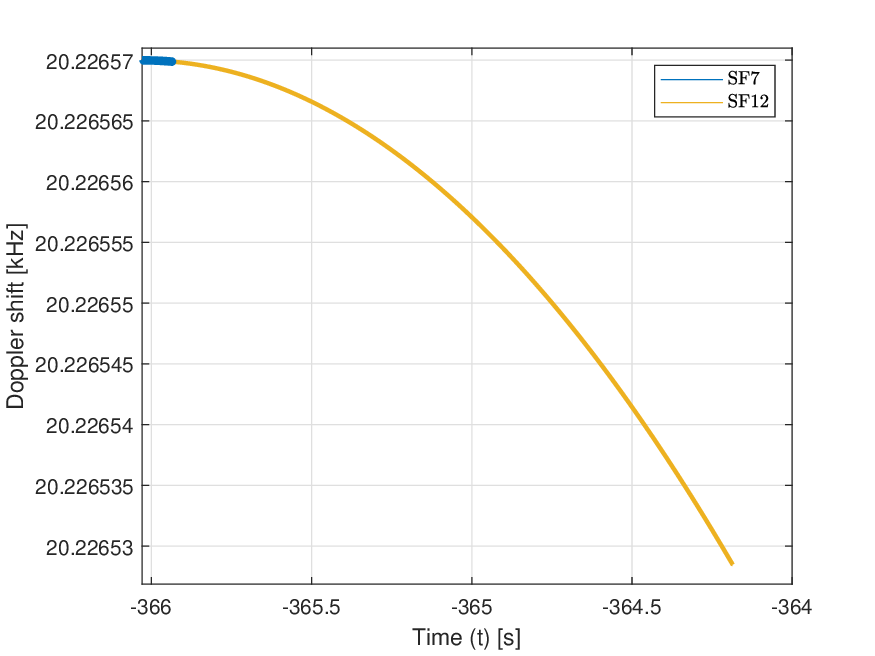}}
    \subfloat[\label{fig:dopplerScenario2} \textit{Case 2}]{%
        \includegraphics[trim= {3 0 30 20}, clip, width=0.5\linewidth]{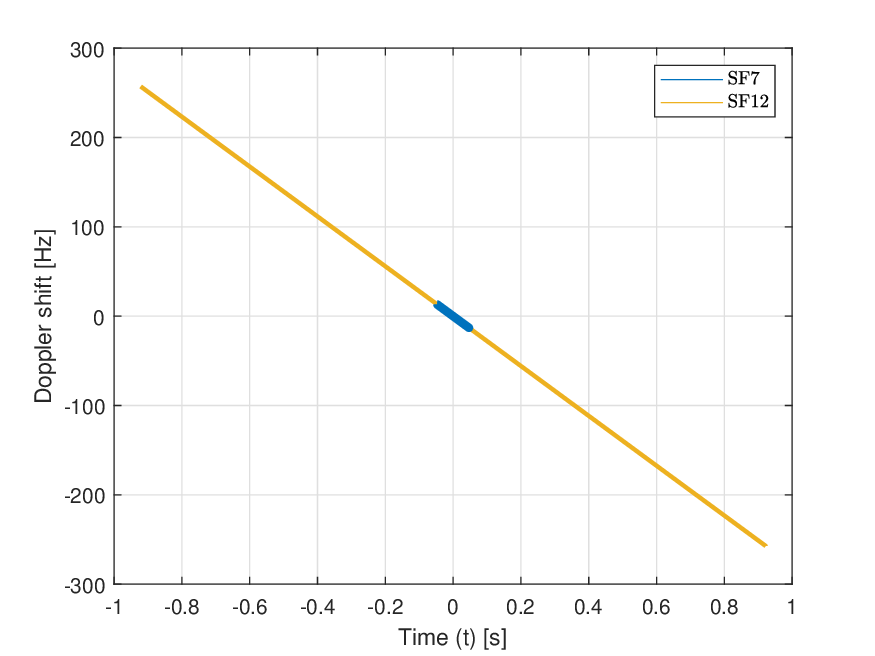}}
    \caption{Doppler shifts are analyzed for different $\mathrm{SF}$ values and satellite positions. \textit{Case 1} is illustrated in Fig.~\ref{fig:dopplerScenario1}, while Fig.~\ref{fig:dopplerScenario2} depict the Doppler effects for \textit{Case 2} with the same $\mathrm{SF}$ values.}
    \label{fig:SF_doppler}
\end{figure}

To highlight the impact of $\mathrm{SF}$ variations under different Doppler conditions, Fig.~\ref{fig:SF_doppler} investigates the Doppler shifts as a function of time for different satellite positions along the orbit and $\mathrm{SF}$ values. As an illustrative example, Fig.~\ref{fig:SF_doppler} shows the Doppler shift expected by the receiver over a large time window (approximately 2 seconds) for the lowest and highest $\mathrm{SF}$ values, namely $\mathrm{SF}=7$ and $\mathrm{SF}=12$. The scenario considers a satellite orbiting at an altitude of $550$~km, a carrier frequency $F_{\mathrm{C}}=868$~MHz, and a payload size of $15$~Bytes. The corresponding bit rates, following~\cite{Maleki.2024}, are given by $R_\mathrm{b} = \frac{\textrm{CR}\,B}{2^{\mathrm{SF}-1}}$, where $\textrm{CR} = 1$ is the coding rate and $B = 125$ kHz is the bandwidth. For $\mathrm{SF} = 7$, this results in a rate of $R_b \approx  6835$~bits/s, while for $\mathrm{SF} = 12$, the rate reduces to $R_b \approx 366$~bits/s. These values reflect the faster and slower chirping rates associated with low and high $\mathrm{SF}$ values, respectively.
\begin{itemize}
    \item \textit{Case 1:} This scenario corresponds to the satellite's position leading to the maximum Doppler shift along with a minima-and negative-Doppler rate. As shown in Fig.~\ref{fig:DopplerEffects}, this situation occurs approximately at times $t = -366$~s. This condition results in a substantial yet slowly varying frequency shift caused by the satellite's relative motion.
    
    \item \textit{Case 2:} This scenario focuses on the satellite positioned as to experience the maximum Doppler rate with minimal Doppler shift. As shown in Fig.~\ref{fig:DopplerEffects}, at $t = 0$~s the Doppler rate is at its highest, although the overall Doppler shift remains relatively low.
\end{itemize}

Specifically, \textit{Case 1} results in higher Doppler shifts (approximately 20 kHz) for both $\mathrm{SF}=7$ and $\mathrm{SF}=12$, but with relatively low Doppler rates. In fact, the Doppler shift remains nearly constant at around 20 kHz throughout the entire frame duration. In contrast, \textit{Case 2} exhibits more pronounced Doppler rate effects, particularly for $\mathrm{SF}=12$, due to its longer \ac{ToA}. Specifically, for $\mathrm{SF}=12$, the Doppler shift varies by approximately 510~Hz over the frame duration, compared to about 25~Hz for $\mathrm{SF}=7$. As anticipated, the extended \ac{ToA} associated with higher $\mathrm{SF}$ values increases susceptibility to Doppler effects, as the receiver is more affected by the rapid frequency variations characteristic of \textit{Case 2}.

To address these challenges, robust estimation and compensation techniques are essential for mitigating the Doppler effect. The following section presents four strategies for Doppler estimation and compensation, specifically designed for \ac{LoRa} \ac{DtS} links in \ac{LEO} satellite scenarios.

\section{Proposed Doppler Estimation Strategies} \label{sec:proposed_frameworks}

In this section, four Doppler estimation strategies are presented, denoted as  \emph{point}, \emph{linear}, \emph{midamble-point} and \emph{midamble-linear} estimation. 

\subsection{Point Estimation}
\label{Sec:point_estimate}

\begin{figure}[!t]  
\centering  
\includegraphics[width=8.8cm]{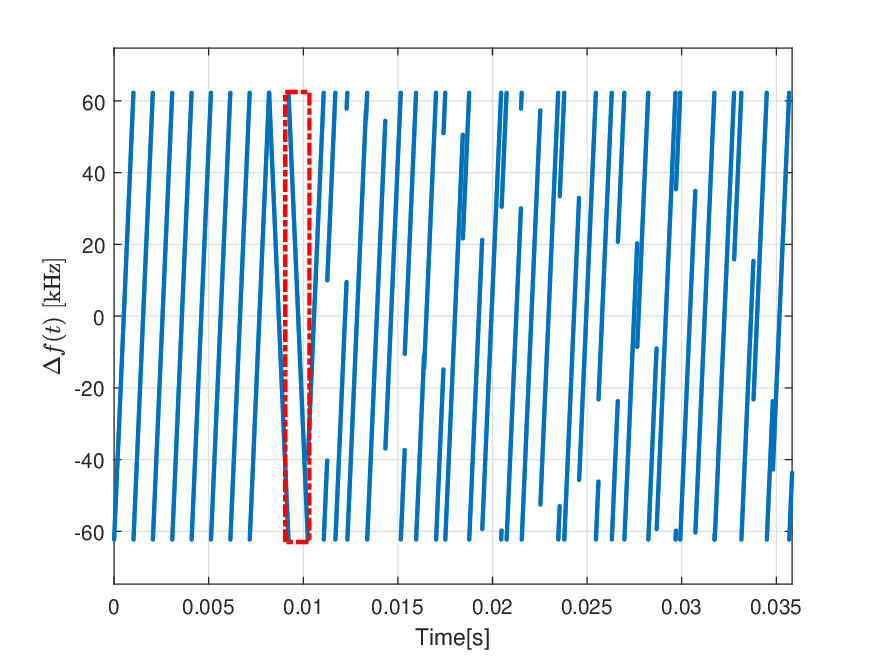}  
\caption{Frequency deviation within a \ac{LoRa} frame with $n_{\text{up}}=8$, $n_{\text{dw}}=2$ and $n_{\text{sym}}=25$. The dotted-dashed red line indicates the last synchronization downchirp used for point Doppler estimation.}  
\label{fig:point_estimation}  
\end{figure}  
\label{Sec.point_estimated}
The point estimation strategy leverages the last synchronization downchirp in the preamble of the \ac{LoRa} frame, illustrated in Fig.~\ref{fig:point_estimation}, which is used to estimate the Doppler effect in the subsequent payload.

Similar to the procedure for demodulating an upchirp corresponding to a symbol $s_n$ from the modulation alphabet, as explained in Sec.~\ref{Sec:LoRa_demodulation}, the synchronization downchirp is demodulated by multiplying its complex envelope by the complex envelope of a pure upchirp, whose frequency increases linearly from $-\frac{B}{2}$ to $ \frac{B}{2}$.
The result of this product, $s_{\text{point}}(t)$, is then subjected to a Fourier transform. The bin $S_{\text{point}}$, where the peak of the Fourier transform appears corresponds to the Doppler shift experienced by the considered synchronization downchirp, so that
\begin{equation}
    S_{\text{point}} = \argmax_{f} \left| \mathcal{F} \{s_{\text{point}}(t) \} \right|,
\end{equation}
with  $\mathcal{F}\{.\}$ denoting the Fourier transform. Clearly, in the absence of any Doppler-induced frequency shift, $S_{\text{point}}=0$.
The estimated Doppler-induced phase shift is then obtained as
\begin{equation}
\label{eq:Theta_point}
    \hat{\theta}_{\mathrm{point}}(t) = 2\pi\,S_\text{point}\,t.
\end{equation}

Given \eqref{eq:Theta_point}, the complex envelope $i_n(t)$ of the received chirp \eqref{eq:dechirped_signal}, with $n$ spanning all chirps in the payload, is multiplied by $\exp\left(-j \, \hat{\theta}_{\mathrm{point}}(t)\right)$ to obtain the Doppler-compensated signal with complex envelope
\begin{equation}  
\label{eq:dechirped_point_compensated}
   \hat{i}_n(t) = i_n(t) \exp\left(-j \, \hat{\theta}_{\mathrm{point}}(t)\right).
\end{equation}

This operation corrects for the Doppler-induced phase variations across the entire payload, assuming that the Doppler shift remains constant throughout the entire \ac{ToA}. The compensated signal, as expressed in \eqref{eq:dechirped_point_compensated}, is then demodulated by performing the dechirping operation \eqref{eq:dechirped_signal}, followed by the Fourier transform, as described in~\cite{Pasolini.2022}.

\begin{remark} 
This strategy is quite simple and straightforward, representing a simplified version of the solution proposed in \cite{sym14040747}, which estimates the Doppler shift as the average shift experienced across all upchirps in the preamble. Here, we decided to leverage only the last downchirp in the preamble because it is the closest to the payload, which is the part of the frame that needs to be demodulated with the highest accuracy. Conversely, the averaging operation carried out in \cite{sym14040747} makes the estimated Doppler shift dependent also on the upchirps that are farther from the payload, increasing the risk that the estimate does not accurately reflect the Doppler conditions affecting the payload. Nevertheless, neither this approach nor the one in \cite{sym14040747} can compensate for the Doppler rate, as the Doppler shift, estimated from the preamble, is considered constant throughout the entire \ac{ToA}. This approach serves as a baseline in this paper to evaluate the improvements achievable with more sophisticated strategies.
\end{remark}

\subsection{Linear Estimation} \label{Sec:Linear_estimation}

\begin{figure}[!t]
\centering
\includegraphics[width=8.8cm]{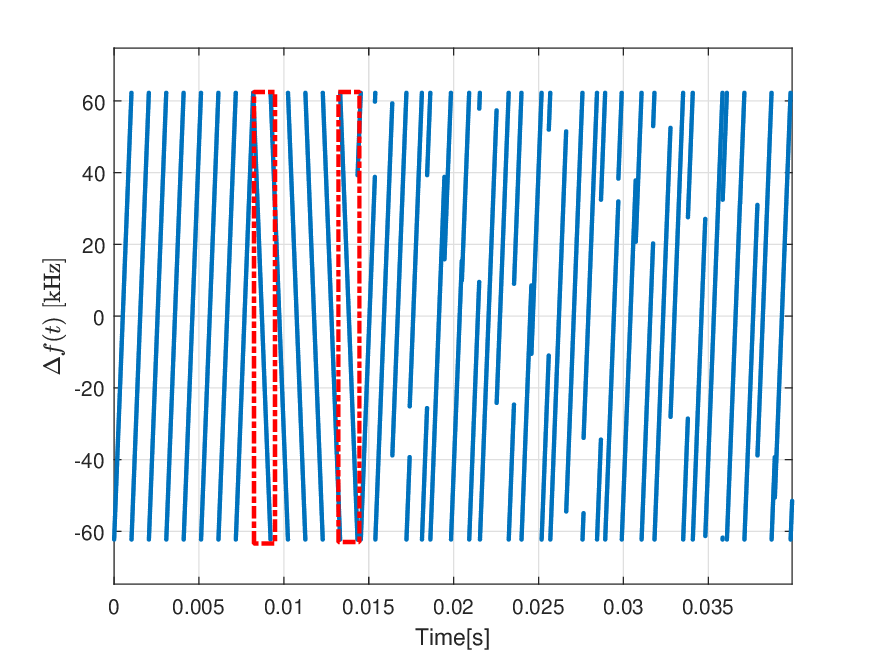}
\caption{Frequency deviation within a \ac{LoRa} frame  with $n_{\text{up}}=8$, $n_{\text{dw}}=6$ and $n_{\text{sym}}=25$. The dotted-dashed red line indicates the two downchirps used for linear Doppler estimation.} 
\label{fig:linear_estimation}
\end{figure}

The linear estimation approach takes into account the first and last synchronization downchirps within a \ac{LoRa} frame, as highlighted in Fig.~\ref{fig:linear_estimation}.

Specifically, the estimated Doppler shift at any given time instant $t$ within the frame duration is obtained using a linear model that interpolates between the frequency deviations at the first and last synchronization downchirps. Therefore, the frequency deviation due to the Doppler effect is modeled as 
\begin{equation}
    \gamma(t) = \alpha (t - T_\text{start}) + S_{\text{linear}}[1],
    \label{eq:linear_freq_estimation}
\end{equation}
where $S_{\text{linear}}[n]$, with $n \in \{1, \dots, n_{\text{dw}}\}$, denotes the frequency bins in which the spectrum peaks appear for each synchronization downchirp. Here, $T_\text{start}$ denotes the time instant corresponding to the beginning of the first downchirp, which experiences the estimated Doppler shift $S_{\text{linear}}[1]$.

In light of the above, the angular coefficient $\alpha$, which quantifies the rate of change of the frequency deviation over time, is computed as
\begin{equation}
    \alpha = \frac{S_{\text{linear}}[n_{\text{dw}}] - S_{\text{linear}}[1]}{T_\text{c} (n_{\mathrm{dw}} - 1)},
\end{equation}
where $S_{\text{linear}}[n_{\text{dw}}]$ and $S_{\text{linear}}[1]$ are obtained using the same procedure described in Sec.~\ref{Sec:point_estimate} to derive $S_{\text{point}}$.

This formulation models the variation of the Doppler shift during the reception of the downchirps as linear and assumes that this linear trend persists throughout the entire frame duration, i.e., for the entire \ac{ToA}. This approach offers a potential improvement over baseline point estimation introduced in Sec.~\ref{Sec.point_estimated}, where a single frequency shift is estimated and applied to compensate for the Doppler effect over the entire frame. This is particularly relevant when the Doppler rate is significant, such as when the satellite is at its closest distance to the \ac{IoT} device, as in Fig.~\ref{fig:dopplerScenario2}.

Given \eqref{eq:linear_freq_estimation}, the estimated Doppler-induced phase shift is immediately obtained as follows
\begin{equation}
    \hat{\theta}_{\mathrm{linear}}(t) = 2\pi\,t \left[ \frac{1}{2}\alpha t - \alpha T_\text{start} + S_{\text{linear}}[1] \right].
    \label{eq:Theta_linear}
\end{equation}
This estimate is then used to obtain the Doppler-compensated signal
\begin{equation}  
    \hat{i}_n(t) = i_n(t) \exp\left(-j \, \hat{\theta}_{\mathrm{linear}}(t)\right),  
\end{equation} 
which is subsequently demodulated, through dechirping and Fourier transform, following the procedure outlined in~\cite{Pasolini.2022}.
 
\begin{remark}
As clearly shown in Fig.~\ref{fig:DopplerEffects}, the Doppler shift variation during the entire satellite visibility period does not follow a linear trend. At first glance, this appears to contradict the fundamental assumption underlying the strategy introduced in this section---namely, that the Doppler shift varies linearly over time. However, the key requirement for this strategy to be effective is that the Doppler shift remains approximately linear within the \ac{ToA} of the received frame. Since the \ac{ToA} depends on various factors, such as the selected SF and payload length, and given that the profile of the Doppler shift within the \ac{ToA} is influenced by the satellite’s orbital position, this aspect requires further investigation, which is addressed in the numerical results section.

It is worth noting, furthermore, that in order to increase the accuracy of $\alpha$, the two downchirps used for its estimation should be well separated. This would require increasing the number of downchirps in the preamble and, consequently, modifying the frame format compared to the one currently defined, which includes only 2.25  downchirps. This is exemplified in Fig.~\ref{fig:linear_estimation}, which shows a frame with six downchirps.
\end{remark}

\subsection{Midamble-based Estimations}

In principle, the linear estimation strategy presented in Sec.~\ref{Sec:Linear_estimation} constitutes a refinement of the point estimation approach, as it incorporates some Doppler rate effects. However, its accuracy depends on the assumption that the Doppler shift varies linearly over the frame's \ac{ToA}. In scenarios where the Doppler shift exhibits non-linear variations over the duration of a frame, as can occur with long frames, this assumption may result in estimation inaccuracies.

\begin{figure}[!t]
\centering
\includegraphics[width=8.8cm]{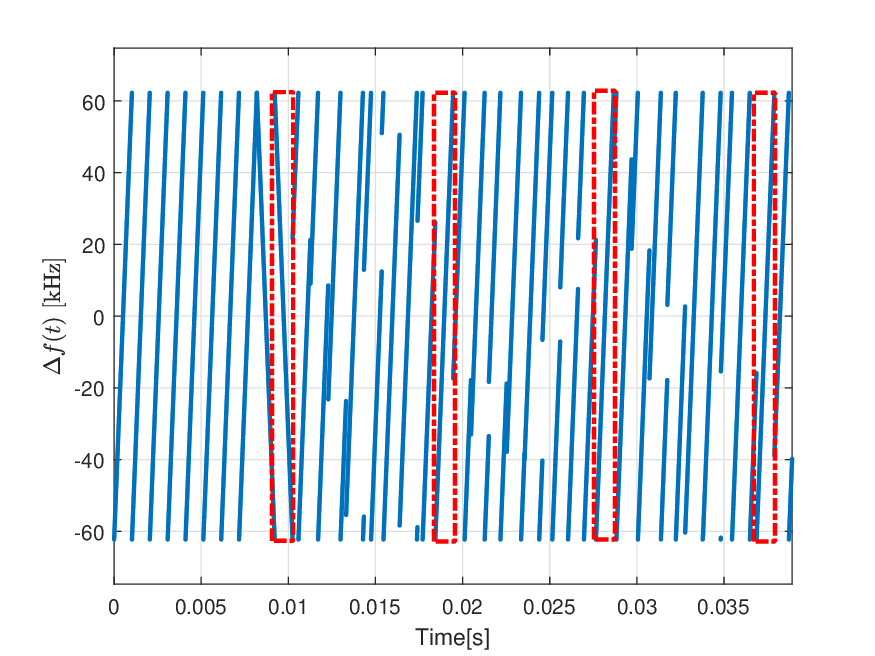}
\caption{Frequency deviation within a \ac{LoRa} frame  with $n_{\text{up}}=8$, $n_{\text{dw}}=2$ and $n_{\text{sym}}=25$. The dotted-dashed red lines indicate the preamble downchirp and the midamble upchirps used for midamble-point Doppler estimation updates.}
\label{fig:midamble_point_estimation}
\end{figure}

\begin{figure}[!t]
\centering
\includegraphics[width=8.8cm]{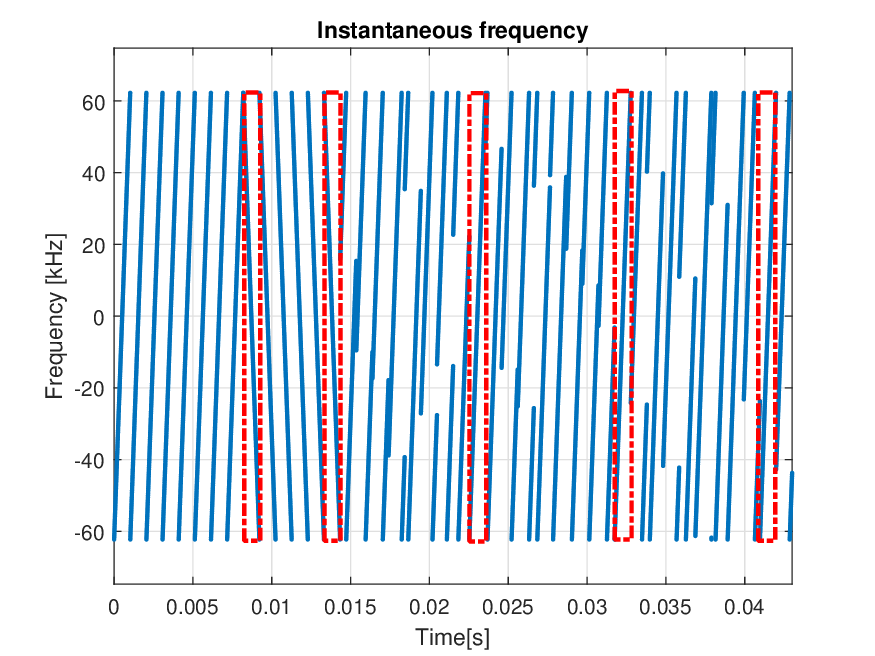}
\caption{Frequency deviation within a \ac{LoRa} frame  with $n_{\text{up}}=8$, $n_{\text{dw}}=6$ and $n_{\text{sym}}=25$. The dotted-dashed red lines indicate the preamble downchirp and the midamble upchirps used for midamble-linear Doppler estimation updates.}
\label{fig:midamble_linear_estimation}
\end{figure}

To address this limitation, the midamble-point and midamble-linear estimation strategies refine the point and linear estimation approaches, respectively. These strategies leverage pilot upchirps embedded within the payload to iteratively refine the Doppler estimate. These pilots, referred to as midambles, are strategically placed at regular intervals throughout the payload, enabling the receiver to dynamically adapt to Doppler variations. 
From a practical standpoint, a midamble consists of a pure upchirp, which corresponds to symbol 0 in the modulation alphabet. Therefore, inserting a midamble into the payload at the transmitter side simply involves placing a 0 in the appropriate position within the symbol sequence representing the transmitted data. These additional symbols are, of course, discarded by the receiver when extracting the actual data.
Examples of the resulting frames for both strategies are illustrated in Fig.~\ref{fig:midamble_point_estimation} (midamble-point strategy) and Fig.~\ref{fig:midamble_linear_estimation} (midamble-linear strategy).

\subsubsection{Midamble-Point Estimation}

The midamble-point strategy begins with an initial estimate of the Doppler-induced phase shift, denoted as $\hat{\theta}_{\text{point,mid}}[1]$, computed using the last downchirp of the preamble, similar to~\eqref{eq:Theta_point}. This estimate is then iteratively refined as new midambles are received. Specifically, the Doppler-induced phase shift at the $m$-th midamble update step, where $m \in \{2, \dots, n_{\text{point,mid}}\}$ and $n_{\text{point,mid}}$ is the total number of midambles, is given by
\begin{equation}
\hat{\theta}_{\text{point,mid}}[m] = \hat{\theta}_{\text{point,mid}}[m-1] + \Delta \hat{\theta}_{\text{point,mid}}[m],
\label{eq:midamble_point_phase_update}
\end{equation}
where $\Delta \hat{\theta}_{\text{point,mid}}[m]$ is the correction term derived from the frequency offset estimated using the $m$-th midamble. This correction term is computed using the same procedure described in Sec.~\ref{Sec:point_estimate} for point estimation, where the peak of the Fourier transform of the dechirped midamble signal corresponds to the Doppler shift~\footnote{Note that the proposed design iteratively refines the Doppler shift estimate using upchirps in the midambles (within the payload), in contrast to the initial estimate derived from downchirps in the preamble.}.

The Doppler-compensated signal for the $n$-th chirp in the payload is then obtained by applying the updated phase shift estimate
\begin{equation}
\hat{i}_n(t) = i_n(t) \exp\left(-j \, \hat{\theta}_{\mathrm{point,mid}}(t)\right).
\label{eq:midamble_point_compensated_signal}
\end{equation}

\begin{remark}
According to this strategy, the Doppler shift is modeled as a piecewise constant function.
The effectiveness of this approach depends on the periodicity of midamble insertion within the payload, which should be determined based on the expected Doppler rate of the satellite link and the system’s tolerance to Doppler-induced frequency variations.

The update interval, $T_{\text{point,mid}}$, for midamble-point-based Doppler compensation, can be approximated as
\begin{equation}
T_{\text{point,mid}} = \frac{k}{|\xi|} \cdot R_s,
\label{eq:midamble_point_update_interval}
\end{equation}
where $k$ is a constant factor representing the system's tolerance to frequency deviations\footnote{Following~\cite{Ameloot.2021}, a frequency shift caused by the Doppler effect results in a mismatch $\Delta f$ between the \ac{LoRa} receiver's frequency settings and the incoming signal. If the receiver is perfectly time-aligned with the \ac{LoRa} symbol and $|\Delta f| > 0.5\,B/M$, a symbol error occurs. Even when this condition is not met, noise may still cause symbol errors. Therefore, we consider typical values of $k$ ranging from 0 to 0.5.}, and $\xi$ is the rate of change of the Doppler shift over the measurement interval, with $R_s = \frac{B}{2^{\mathrm{SF}}}$ denoting the symbol rate. The optimal number of midambles $n_{\text{point,mid}}^\star$ is then given by
\begin{equation}
n_{\text{point,mid}}^\star = \left\lceil \frac{T_\text{c}\,n_{\text{sym}}}{T_{\text{point,mid}}} \right\rceil = \left\lceil \frac{n_{\text{sym}}}{n_{\text{int, point}}} \right\rceil,
\label{eq:optimal_midambles}
\end{equation}
where $n_\text{int,point}$ is the number of chirps between consecutive midambles for midamble-point strategy. For instance, consider $B=125$~kHz, $\mathrm{SF}=10$, $k=0.1$, $n_{\text{sym}}=15$, ${T_\text{c}=8.2\cdot10^{-3}}$~s and $\xi=-304.71$~Hz/s. Substituting these values into~\eqref{eq:midamble_point_update_interval} yields $T_{\text{point,mid}}=0.0401$~s. Applying this result to~\eqref{eq:optimal_midambles} yields $n_{\text{point,mid}}^\star=4$ midambles.

However, in practice, determining the exact number of midambles required for Doppler compensation on a per-frame basis is impractical, as $\xi$ depends on the relative position between the satellite and the terrestrial device at the time of transmission. Accurately estimating this would require the transmitter to know both its own position and that of the receiver. In the uplink, this would necessitate equipping terrestrial devices with positioning capabilities (e.g., GNSS) and access to satellite ephemerides, while in the downlink, the satellite would need precise knowledge of the receiver’s coordinates. Moreover, a computational effort is required for each transmission. A more practical approach is to adopt a fixed-interval midamble insertion strategy, where the number and placement of midambles are predetermined based on the system design and expected worst-case operating conditions. While this method simplifies implementation, it may lead to suboptimal performance when the Doppler shift varies significantly over the frame duration. The trade-offs associated with the number of midambles are further explored in the numerical results section.
\end{remark}

\subsubsection{Midamble-Linear Estimation}
The midamble-linear strategy extends the basic linear estimation approach described in Sec.~\ref{Sec:Linear_estimation} by dynamically refining the Doppler shift estimate using midambles distributed throughout the frame. Unlike the static linear model, this method iteratively updates both the Doppler shift and its rate of change, allowing for  tracking of time-varying Doppler effects within the payload as well.

The process begins with an initial estimate of the Doppler-induced phase shift, obtained from the preamble downchirps using the procedure outlined in Sec.~\ref{Sec:Linear_estimation}. 
This provides a baseline Doppler shift $S_{\text{linear,mid}}[1]$ and rate $\alpha_{\mathrm{mid}}[1]$, where $m=1$ refers to the initial estimates obtained from the preamble downchirps. 

For each subsequent midamble $m \in \{2, \dots, n_{\mathrm{linear,mid}}\}$ in the payload, with $n_\text{int,linear}$ being the number of chirps between consecutive midambles, the Doppler shift $S_{\text{linear,mid}}[m]$ is estimated by locating the peak of the Fourier transform of the dechirped midamble signal, following the same procedure as in Sec.~\ref{Sec:point_estimate}. The updated Doppler rate $\alpha_{\mathrm{mid}}[m]$ is then computed as the slope between the current and previous midamble, given by 
\begin{equation}
\alpha_{\mathrm{mid}}[m] = \frac{S_{\text{linear,mid}}[m] - S_{\text{linear,mid}}[m-1]}{T_m - T_{m-1}},
\end{equation}
where $T_m$ and $T_{m-1}$ denote the time instants of the $m$-th and $(m-1)$-th midambles, respectively. This refined rate is used to adjust the phase shift estimate for the next segment, expressed as

\begin{equation}
    \begin{split}
    \hat{\theta}_{\mathrm{linear,mid}}(t) = 2\pi t \bigg[ &\frac{1}{2}\alpha_{\text{mid}}[m] t -  \alpha_{\text{mid}}[m] T_{m-1} \\
    &+ S_{\text{linear,mid}}[m-1] \bigg].
    \end{split}
\end{equation}

The compensated signal for each payload chirp is then obtained by applying the corresponding phase correction
\begin{equation}  
\hat{i}_n(t) = i_n(t) \exp\left(-j \hat{\theta}_{\mathrm{linear,mid}}(t)\right).
\end{equation}

\begin{remark}
According to this strategy, the Doppler shift is modeled as a piecewise linear function, allowing for varying slopes in the intervals between midambles. This approach potentially provides superior tracking of Doppler variations compared to the basic linear estimation, particularly for longer frames where non-linear effects become significant. The trade-off between estimation accuracy and additional midambles introduced will be quantitatively analyzed in the numerical results section.
\end{remark}

\section{Numerical Results}
\label{sec:numerical_results}

This section presents numerical results evaluating the performance of the proposed Doppler estimation and compensation strategies. The assessment was conducted using a \ac{LoRa} physical-layer simulator, validated in~\cite{Pasolini.2022}, to model a satellite communication link. The simulated link consists of a \ac{LoRa} transmitter, a channel affected by \ac{AWGN} and Doppler effect, and a receiver. The model 
 of the satellite link is based on the analysis provided in~\cite{Hourani.2024}, which investigates the Doppler shift effect in \ac{LEO} satellite constellations. That work elucidates the analytical bounds of the Doppler shift, highlighting its dependence on the orbital height. These bounds enable accurate modeling of satellite trajectories and the corresponding Doppler shifts, ensuring a realistic representation of the communication channel in the simulator.

\subsection{Scenarios, Parameters and Performance Metric}

Various scenarios were considered, accounting for different satellite positions and multiple \ac{LoRa} configurations, as detailed in Table \ref{tab:scenarios_parameters}. Specifically, we analyzed the uplink of a \ac{DtS} \ac{LoRa} link established by an \ac{IoT} terrestrial device communicating with a \ac{LEO} satellite at an altitude of $550$~km. At the maximum elevation angle, the satellite was assumed to be directly overhead of the \ac{IoT} device. Two satellite positions along the orbit, as illustrated in Fig.~\ref{fig:SF_doppler}, were examined: \textit{Case~1}, characterized by a high Doppler shift and a low Doppler rate (corresponding to a low elevation angle), and \textit{Case~2}, exhibiting the opposite conditions (corresponding to a high elevation angle).
\begin{table}[t]
    \centering
    \caption{Scenarios and Parameter Settings}
    \label{tab:scenarios_parameters}
   \resizebox{\columnwidth}{!}{ \begin{tabular}{l|l}
         \toprule
         \textbf{Orbital Height} & 550~km \\ \midrule
         \textbf{Satellite Position} & \textit{Case 1}, \textit{Case 2} \\ \midrule
         \textbf{Bandwidth} & 125~kHz \\ \midrule
         \textbf{Carrier Frequency} ($\bm{F}_{\mathbf{C}}$) & 868~MHz \\ \midrule
         \textbf{Spreading Factors (SFs)} & 7, 10, 12 \\ \midrule
         \textbf{LDRO} & On, Off \\ \midrule
         $\bm{n}_{\mathbf{up}}$ & 8 \\ \midrule
         $\bm{n}_{\mathbf{dw}}$ & 2, 6 \\ \midrule
         $\bm{n}_{\mathbf{int,point}}$ & $\mathrm{SF=12}:\,\, 1$, $\mathrm{SF=10}:\,\, 4$, others: $12$ \\ \midrule
         $\bm{n}_{\mathbf{int,linear}}$ & $6$ \\ \midrule
         \textbf{Payload Size} & 120~bits \\ \midrule
         \textbf{Coding Rate (CR)} & 1 \\  
         \bottomrule
    \end{tabular}}
\end{table}

Regarding the \ac{LoRa} signal, we considered a bandwidth of ${B = 125}$~kHz, centered at a carrier frequency of ${F_\mathrm{C} = 868}$~MHz, with spreading factors $\mathrm{SF}=7$, $\mathrm{SF}=10$, and $\mathrm{SF}=12$. Additionally, we investigated the impact of the \ac{LDRO} mode, which enhances robustness against Doppler effects.

For the frame format, the preamble consisted of $n_{\text{up}}=8$ upchirps, followed by $n_{\text{dw}}=2$ downchirps for the point-based estimation strategies, or $n_{\text{dw}}=6$ downchirps for the linear-based estimation strategies. The choice of $n_{\text{dw}}=2$ for the point-based strategies is based on the standard \ac{LoRa} frame structure, which includes two complete downchirps. For the linear-based estimation strategies, $n_{\text{dw}}=6$ is chosen to improve the estimation of the angular coefficient $\alpha$, which represents the rate of change of the frequency deviation over time, by utilizing a larger number of downchirps in the preamble.  Additionally, unless stated otherwise, the payload size was set to $L=120$~bits. In all cases, we assumed that the parameter CR, which denotes the coding rate according to \ac{LoRa} specification, is set to 1, meaning that one parity bit is added for every four data bits. For the midamble-based strategies, the interval $n_{\text{int,point}}$ between consecutive midambles in the midamble-point approach is set to one chirp at $\mathrm{SF}=12$, four chirps at $\mathrm{SF}=10$, and twelve chirps for $\mathrm{SF}=7$. In the midamble-linear strategy, the interval $n_{\text{int,linear}}$ is set to 6 chirps for all configurations.

For each scenario and parameter setting, the performance of the estimation strategies was evaluated in terms of \ac{SER}---which also corresponds to the chirp error rate.  In addition, we define the \ac{SNR}, as $\text{SNR}=\frac{P}{P_\text{n}}$, where $P$ represents the average received signal power and $P_\text{n}$ the noise power within the bandwidth $B$. Finally, for comparison, performance was also assessed in a benchmark scenario without Doppler effects, considering only \ac{AWGN}.

\subsection{Influence of SNR and Satellite Position}

\begin{figure*}
    \centering
    \setkeys{Gin}{width=0.33\linewidth}
    \subfloat[$\mathrm{SF}=7$\label{fig:SF7_case1}]{\includegraphics{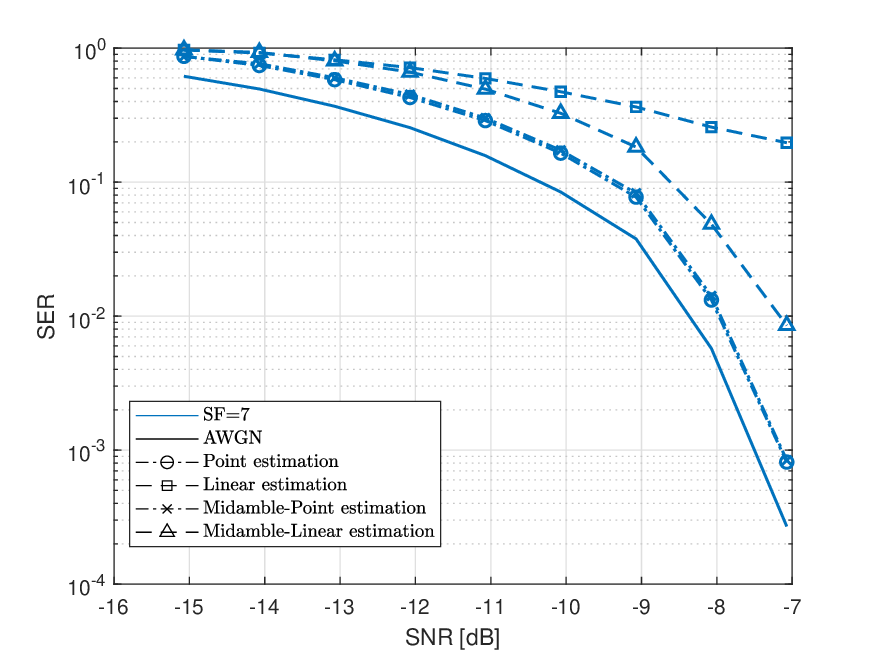}}
    \hfill
    \subfloat[$\mathrm{SF}=10$\label{fig:SF10_case1}]{\includegraphics{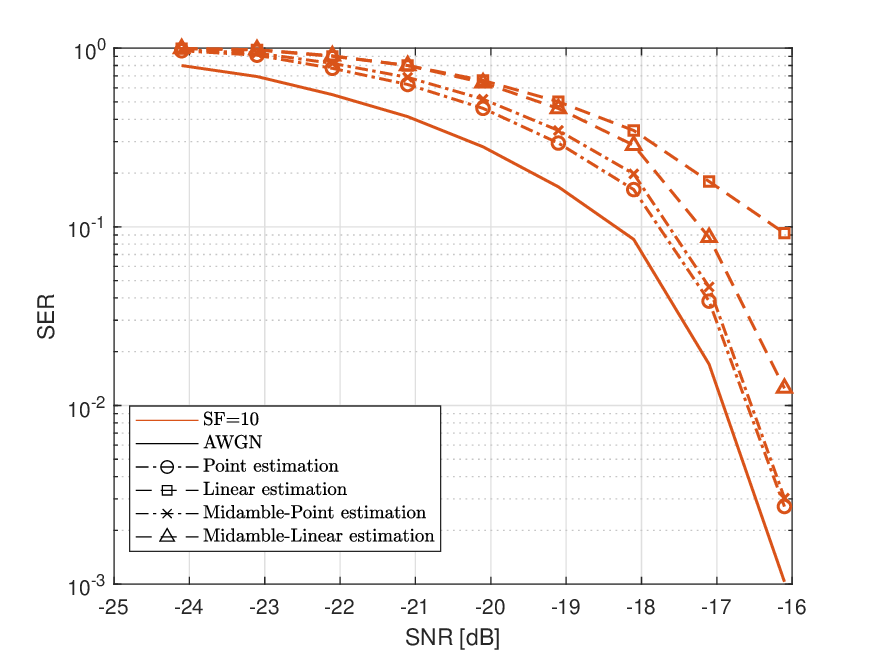}}
    \hfill
    \subfloat[$\mathrm{SF}=12$\label{fig:SF12_case1}]{\includegraphics{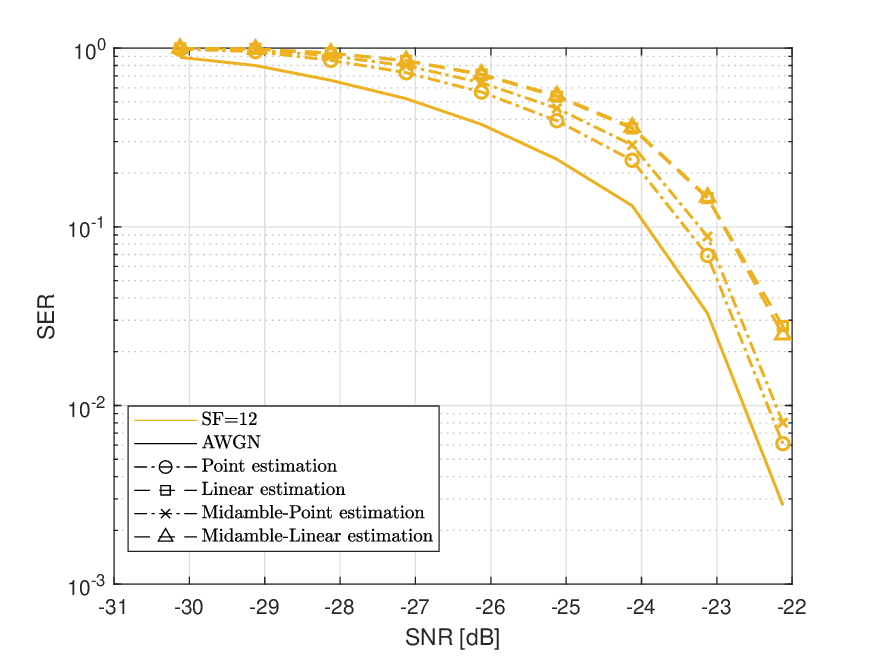}}
    \caption{Symbol error rate for the proposed Doppler estimation strategies and compensation strategies in the \textit{Case~1} scenario.}
    \label{fig:SER_case1}
\end{figure*}

Initially, Fig.~\ref{fig:SER_case1} presents the \ac{SER} as a function of the \ac{SNR} for different spreading factors, $\mathrm{SF} \in \{7, 10, 12\}$, in the \textit{Case 1} scenario, where high Doppler shift and low Doppler rate conditions are considered. As observed, both point and midamble-point estimation techniques yield satisfactory results in terms of error rate performance. Interestingly, the linear-based estimation approaches demonstrate inferior performance, likely due to its assumption of a linear frequency variation over the frame. This assumption fails to accurately model the frequency deviations induced by a high Doppler shift with a very small Doppler rate. Specifically, the Doppler shift in this scenario is not a linear function of time, and by enforcing a linear estimation, we inadvertently amplify the effects of noise and the finite resolution of the FFT in the Doppler shift estimate. However, the midamble-linear approach mitigates this issue to some extent.

\begin{figure*}
    \centering
    \setkeys{Gin}{width=0.33\linewidth}
    \subfloat[$\mathrm{SF}=7$\label{fig:SF7_case2}]{\includegraphics{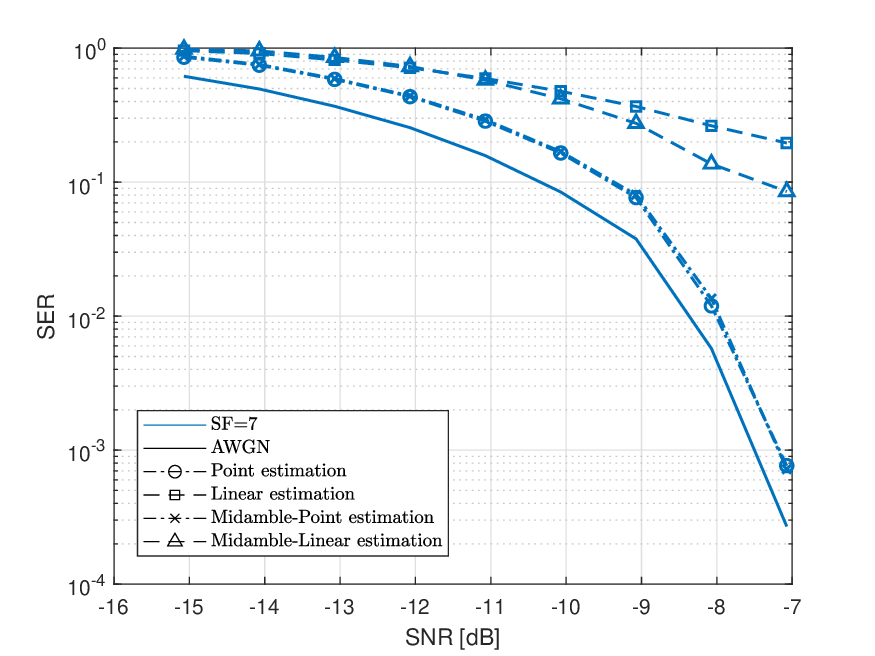}}
    \hfill
    \subfloat[$\mathrm{SF}=10$\label{fig:SF10_case2}]{\includegraphics{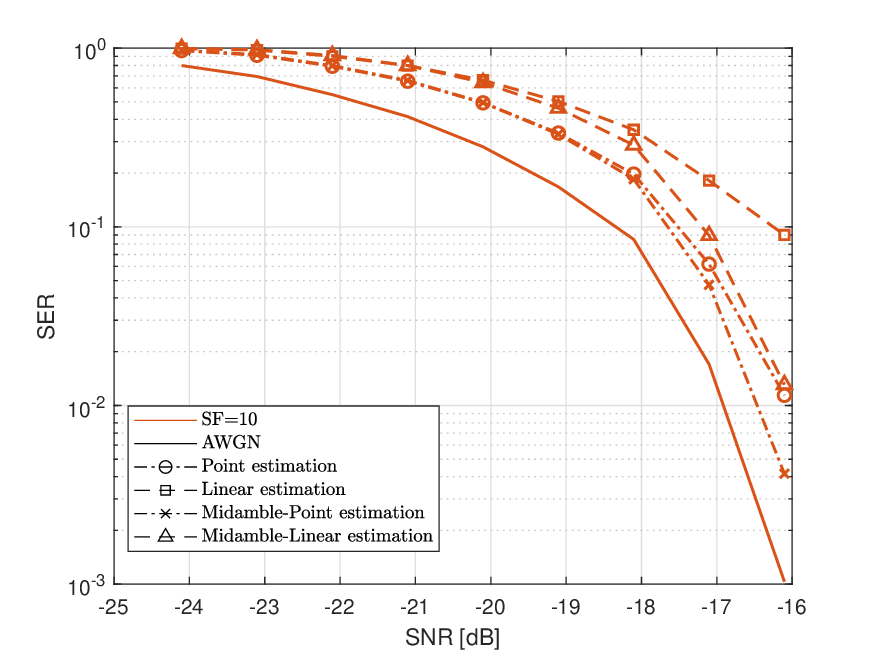}}
    \hfill
    \subfloat[$\mathrm{SF}=12$\label{fig:SF12_case2}]{\includegraphics{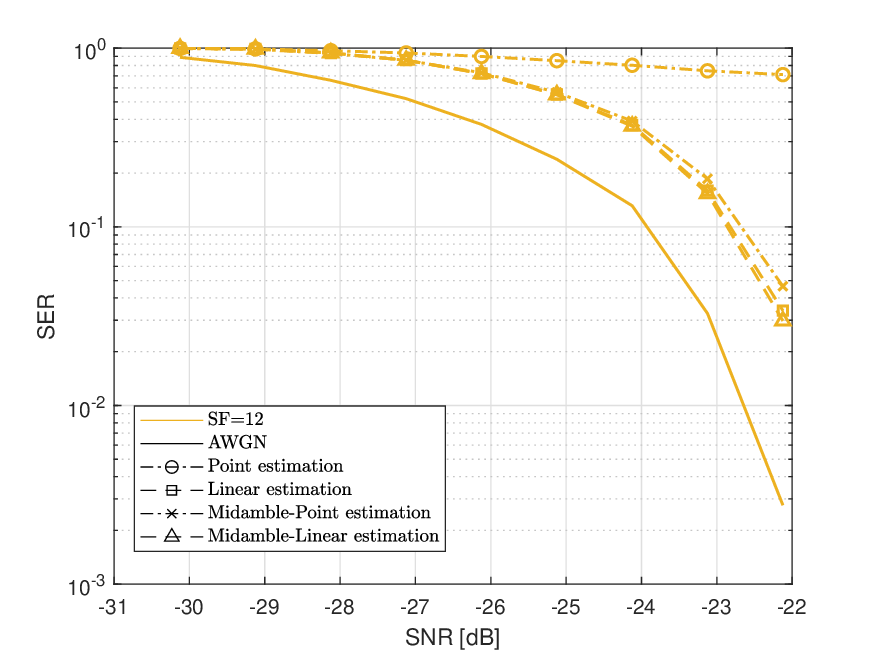}}
    \caption{Symbol error rate for different Doppler estimation and compensation strategies in the \textit{Case~2} scenario.}
    \label{fig:SER_case2}
\end{figure*}

Next, we examine \textit{Case 2}, which considers a scenario with a low Doppler shift but a high Doppler rate. The corresponding error rate results are illustrated in Fig.~\ref{fig:SER_case2}. In this case, the performance of the point estimation method deteriorates as the spreading factor increases. This degradation is attributed to the increased \ac{ToA}, which reduces the precision of the estimation process for higher $\mathrm{SF}$ values. This effect is particularly noticeable for $\mathrm{SF}=12$, where the linear estimation method surpasses the point estimation technique. Furthermore, for both \textit{Case 1} and \textit{Case 2}, we observe that the midamble-point-based approaches consistently deliver performance close to the best technique in each scenario.

\begin{figure}[!t]
    \includegraphics[width=1\linewidth]{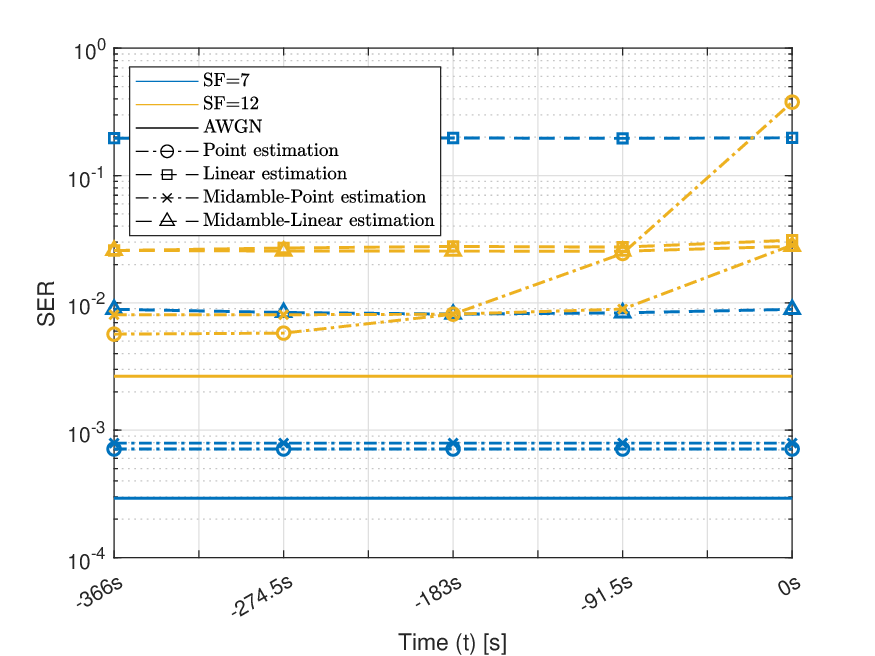}
    \caption{SER performance of different estimation strategies at a fixed $E_\text{s}/N_\text{0}=14$~dB, considering intermediate satellite positions between \textit{Case~1} and \textit{Case~2}.}
     \label{fig:SER_intermediate_cases}
\end{figure}

To further refine the previous analysis, Fig.~\ref{fig:SER_intermediate_cases} presents the \ac{SER} performance of various estimation strategies at a fixed $E_\text{s}/N_\text{0} = 14$~dB. Here, $E_\text{s}$ denotes the energy of the received chirp, while $N_\text{0}$ represents the single-sided noise power spectral density. The parameter $E_\text{s}/N_\text{0}$ characterizes the ratio of symbol energy to noise power spectral density, which is related to the signal-to-noise ratio (SNR) through the expression\footnote{From $\text{SNR}=\frac{P}{P_\text{n}}$, it follows that ${\text{SNR}=\frac{P}{P_\text{n}}=\frac{E_\text{s}/T_\text{c}}{N_\text{0}B}=\frac{E_\text{s}}{N_\text{0}}\frac{1}{BT_\text{c}}}$. Given the relationship $B\,T_\text{c} = 2^{\mathrm{SF}}$, which holds for the \ac{LoRa} modulation, we finally obtain $\text{SNR}=\frac{P}{P_\text{n}}=\frac{E_\text{s}}{N_\text{0}}\frac{1}{2^{\mathrm{SF}}}$.} $\text{SNR} = \frac{E_\text{s}}{N_\text{0}} \frac{1}{2^{\mathrm{SF}}}$. The analysis considers intermediate satellite positions between \textit{Case 1} and \textit{Case 2}, as illustrated in Fig.~\ref{fig:DopplerEffects}.  Specifically, simulations were conducted at time instances $t \in \{-366\text{s}, -274.5\text{s}, -183\text{s}, -91.5\text{s}, 0\text{s}\}$, corresponding to distinct Doppler realizations that progressively transition between the two extreme cases analyzed previously.
For conciseness, we focus on $\mathrm{SF}=7$ and $\mathrm{SF}=12$, representing the two boundary conditions in terms of spreading factor. The results highlight that the midamble-point estimation consistently achieves performance close to the optimal strategy across all intermediate satellite positions, reinforcing its robustness under varying Doppler conditions. Furthermore, the analysis reveals critical transition points, at specific values of $t$, where the performance of the point-based and linear-based estimation techniques intersect, indicating shifts in their relative effectiveness due to the dynamic Doppler characteristics.

\begin{figure}[!t]
    \centering
    \includegraphics[width=1\linewidth]{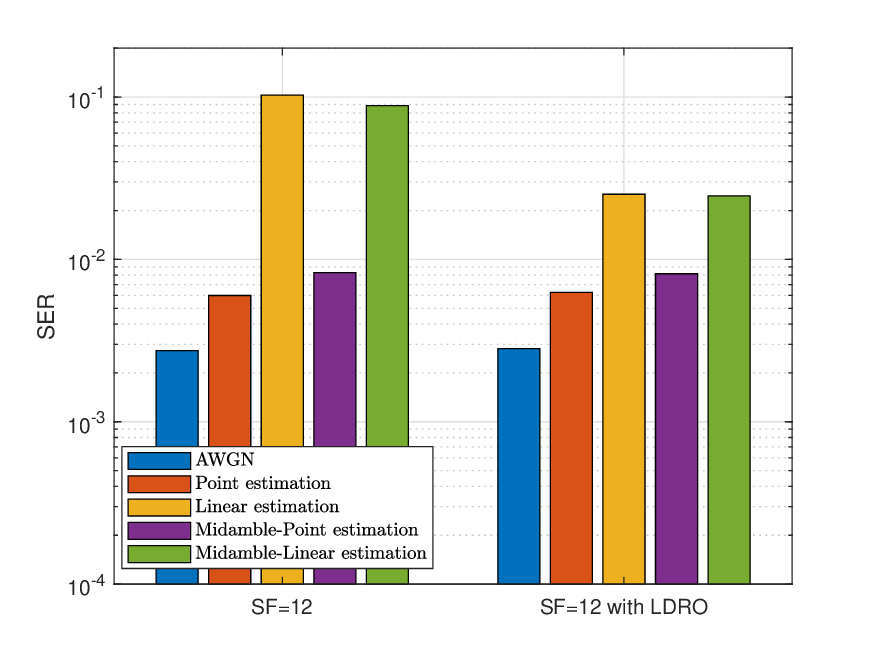}
    \caption{SER performance for $\mathrm{SF}=12$ with and without LDRO, considering fixed $E_\text{s}/N_\text{0} = 14$~dB and $L=120$~bits in the \textit{Case 1} scenario.}
    \label{fig:SER_LDRO_case1}
\end{figure}

\begin{figure}[!t]
    \centering
    \includegraphics[width=1\linewidth]{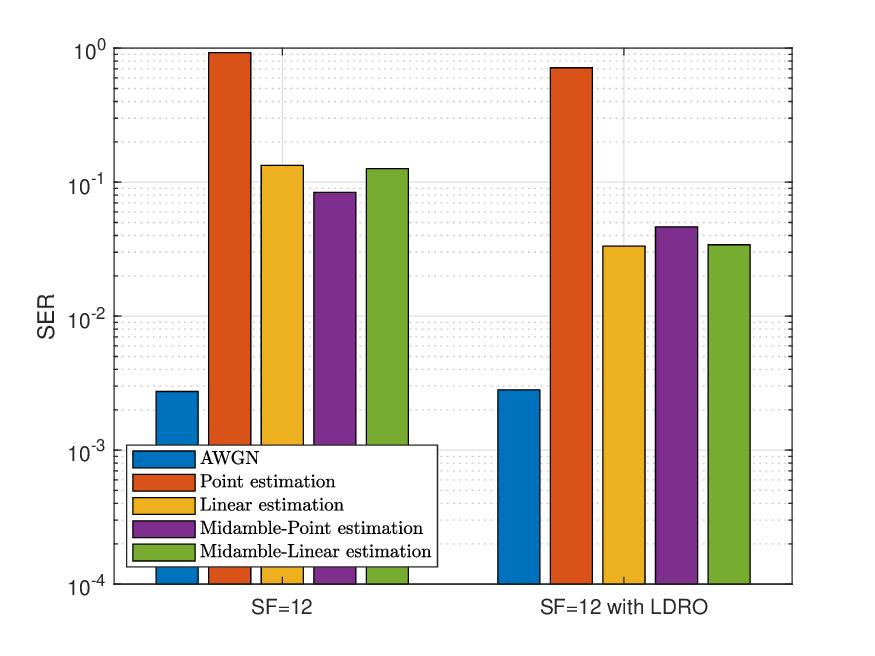}
    \caption{SER performance for $\mathrm{SF}=12$ with and without LDRO, considering fixed $E_\text{s}/N_\text{0}=14$~dB and $L=120$~bits in the \textit{Case 2} scenario.}
    \label{fig:SER_LDRO_case2}
\end{figure}

Finally, Fig.~\ref{fig:SER_LDRO_case1} and Fig.~\ref{fig:SER_LDRO_case2} present a comparative analysis of the \ac{SER} performance for $\mathrm{SF}=12$, evaluating the impact of enabling the \ac{LDRO} mode. The results are obtained under fixed conditions of $E_\text{s}/N_\text{0}=14$~dB and a payload length of ${L=120}$~bits, allowing for a direct assessment of the performance gains achieved through this optimization.  
Fig.~\ref{fig:SER_LDRO_case1} illustrates the results for \textit{Case 1}, while Fig.~\ref{fig:SER_LDRO_case2} depicts the performance for \textit{Case 2}. The findings highlight that enabling \ac{LDRO} improves robustness against Doppler-induced distortions, particularly for \textit{Case 2}, with higher frequency variations.

\subsection{Influence of the Payload Length}

\begin{figure}[!t] 
    \centering \includegraphics[width=1\linewidth]{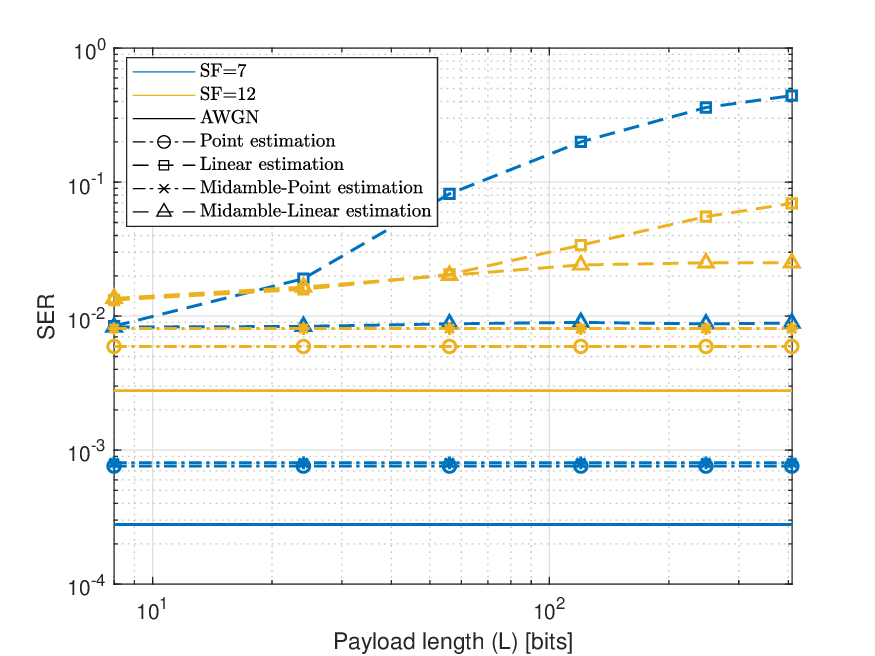} \caption{SER performance as a function of payload length for a fixed $E_\text{s}/N_\text{0} = 14$~dB in the \textit{Case 1} scenario.} 
    \label{fig:SER_payload_case1} 
\end{figure}

\begin{figure}[!t] 
    \centering \includegraphics[width=1\linewidth]{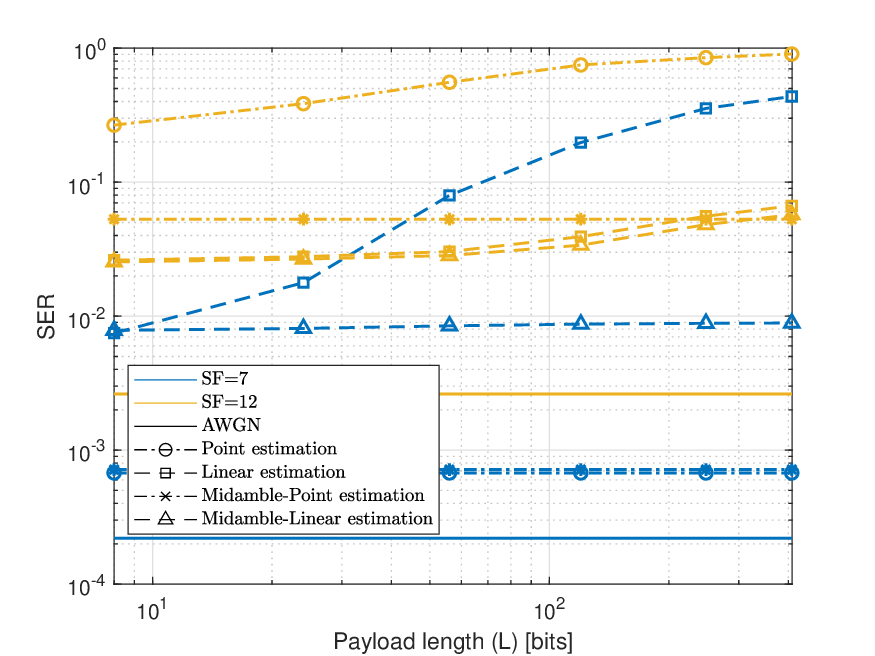} \caption{SER performance as a function of payload length for a fixed $E_\text{s}/N_\text{0} = 14$~dB in the \textit{Case 2} scenario.} 
    \label{fig:SER_payload_case2} 
\end{figure}

The influence of payload length on each estimation framework is illustrated in Fig.~\ref{fig:SER_payload_case1} and Fig.~\ref{fig:SER_payload_case2} for \textit{Case~1} and \textit{Case~2}, respectively. Considering the \ac{LoRaWAN} maximum application payload length for the EU863-870 regional parameters, the maximum payload length is 51 bytes for $\mathrm{SF}=12$, and 222 bytes for $\mathrm{SF}=7$, all operating at a bandwidth of 125 kHz. Therefore, for comparison purposes, we consider a maximum payload size of 51 bytes for all curves, and we set $E_\text{s}/N_0 = 14$~dB. Interestingly, Fig.~\ref{fig:SER_payload_case1} demonstrates that in scenarios with high Doppler shift and low Doppler rate, the impact of payload length is limited to the linear-based estimation frameworks. In this case, the point and midamble-point estimation frameworks perform robustly across Doppler realizations, maintaining high accuracy despite variations in payload length.

However, such stable behavior of the point estimation does not hold when the payload length increases in \textit{Case 2}. In this scenario, due to the high Doppler rate, the estimations provided by both the point and linear estimation frameworks become increasingly inaccurate as the frame's \ac{ToA} increases. This results in a significant rise in the \ac{SER} as $L$ increases. In contrast, the midamble-point strategy, which uses midamble pilot downchirps embedded within the payload to continuously refine the Doppler estimate, offers consistent performance even with longer payloads.

\subsection{Influence of $n_{\text{int}}$}

\begin{figure}[!t]
    \centering
    \includegraphics[width=1\linewidth]{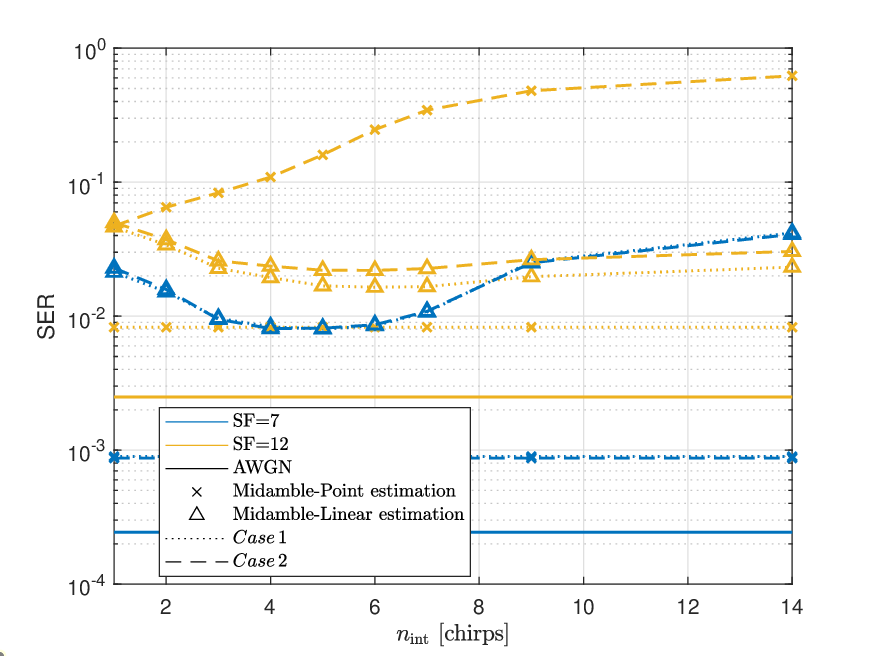}
    \caption{SER as a function of $n_{\text{int}}$ for point and linear midamble estimation frameworks in {\it Case 1} and {\it Case 2} scenarios.}
    \label{fig:SER_nmid}
\end{figure}

To evaluate the impact of $n_{\text{int}}$ on the midamble-point and midamble-linear estimation frameworks\footnote{Hereafter, the terms ``linear'' and ``point'' are omitted when referring to $n_{\text{int}}$ for brevity.}, we assess the \ac{SER} across different spreading factors, considering various midamble insertion intervals within the payload. For this analysis, we also set $E_\text{s}/N_0 = 14$~dB. In this regard, Fig.\ref{fig:SER_nmid} shows how increasing $n_{\text{int}}$ affects estimation accuracy and, consequently, the resulting \ac{SER} in both frameworks.

In particular, the results demonstrate that higher values of $n_{\text{int}}$ are sufficient to achieve optimal midamble-point estimation in scenarios with high Doppler shifts but low Doppler rates (\textit{Case 1}). This observation aligns with prior analyses, as the Doppler effects are often adequately captured by the last downchirp in the preamble of a \ac{LoRa} frame, requiring no further refinement. Conversely, \textit{Case 2} reveals that selecting an appropriate $n_{\text{int}}$ becomes critical in scenarios with both high $\mathrm{SF}$ and high Doppler rates. Here, refining Doppler estimation via the midamble-point strategy yields significant SER improvements. These findings are consistent with the $n_{\text{int}}$ values listed in Table~\ref{tab:scenarios_parameters}. Regarding the midamble-linear estimation, we observe that lower $n_{\text{int}}$ values introduce inaccuracies in the linear estimation. This occurs because closely spaced midamble chirps, analogous to the linear strategy, compromise the precision of $\alpha_{\text{mid}}$. To enhance accuracy, the midambles used for estimation should be sufficiently separated.

\section{Conclusion} \label{sec:final_comments}

This paper investigates the impact of the Doppler effect on \ac{LoRa}-based \ac{DtS} communication in LEO satellite scenarios. In this context, we analyzed both Doppler shift and Doppler rate, proposing and evaluating four distinct Doppler estimation and compensation frameworks: point, linear, midamble-point and midamble-linear estimation. Through comprehensive numerical simulations aligned with \ac{LoRaWAN} specifications, including the \ac{LDRO} feature, we assess the effectiveness of these frameworks in mitigating Doppler-induced distortions and ensuring reliable communication. Our results highlight the critical importance of selecting appropriate estimation strategies based on specific \ac{LoRa} parameters, such as the spreading factor and the interplay between Doppler shift and signal configuration. Among the proposed methods, point-midamble estimation emerges as the most consistent approach, achieving near-optimal accuracy by iteratively refining Doppler shift estimates through midamble symbol integration in \ac{LoRa} frames. In contrast, point and linear estimation offer simpler, lower-complexity alternatives but suffer from reduced accuracy in dynamic Doppler scenarios. Linear-midamble estimation improves upon linear estimation, yet it still underperforms relative to point-midamble estimation in most scenarios.

\bibliographystyle{IEEEtran}
\balance
\bibliography{IEEEabrv,references}

\end{document}